\documentclass[english,prd,superscriptaddress,nofootinbib,preprintnumbers]{revtex4}
\usepackage[latin1]{inputenc}
\usepackage{graphicx}
\usepackage{amssymb}
\usepackage{bm}
\usepackage{color}

\usepackage{amsfonts}
\usepackage{dcolumn}

\def\be{\begin{equation}}
\def\ee{\end{equation}}
\def\ba{\begin{eqnarray}}
\def\ea{\end{eqnarray}}
\def\bs{\begin{subequations}}
\def\es{\end{subequations}}

\newcommand{\A}{{\cal A}}

\usepackage{color}

\makeatother

\usepackage{babel}
\makeatother
\begin{document}

\title{Potential-driven Galileon inflation}

\author{Junko Ohashi}
\affiliation{Department of Physics, Faculty of Science, Tokyo University of Science, 
1-3, Kagurazaka, Shinjuku-ku, Tokyo 162-8601, Japan}

\author{Shinji Tsujikawa}
\affiliation{Department of Physics, Faculty of Science, Tokyo University of Science, 
1-3, Kagurazaka, Shinjuku-ku, Tokyo 162-8601, Japan}

\begin{abstract}

For the models of inflation driven by the potential energy of an inflaton field $\phi$, 
the covariant Galileon Lagrangian $(\partial\phi)^2\Box \phi$ 
generally works to slow down the evolution of the field.
On the other hand, if the Galileon self-interaction is dominant relative to the 
standard kinetic term, we show that 
there is no oscillatory regime of inflaton after the end of inflation.
This is typically accompanied by the appearance of the negative propagation 
speed squared $c_s^2$ of a scalar mode, which leads to the instability of 
small-scale perturbations.
For chaotic inflation and natural inflation we clarify the parameter space 
in which inflaton oscillates coherently during reheating.
Using the WMAP constraints of the scalar spectral index and 
the tensor-to-scalar ratio as well, we find that the self coupling $\lambda$
of the potential $V(\phi)=\lambda \phi^4/4$ is constrained to be very much 
smaller than 1 and that the symmetry breaking scale $f$ of natural inflation
cannot be less than the reduced Planck mass $M_{\rm pl}$.
We also show that, in the presence of other covariant Galileon Lagrangians,
there are some cases in which inflaton oscillates coherently even for
the self coupling $\lambda$ of the order of $0.1$, but still the 
instability associated with negative $c_s^2$ is generally present.

\end{abstract}

\date{\today}


\maketitle

\section{Introduction}

The idea of inflation was originally proposed to address a number of cosmological 
problems plagued in standard Big Bang cosmology \cite{infpapers}.
Moreover inflation provides a causal mechanism for the generation of large-scale 
density perturbations from the quantum fluctuation of 
a scalar field (``inflaton'') \cite{infper}.
The resulting power spectra of scalar and tensor perturbations are 
nearly scale-invariant, whose prediction is consistent with the Cosmic 
Microwave Background (CMB) temperature anisotropies observed by 
COBE \cite{COBE} and WMAP \cite{WMAP1}.

Most models of inflation are based on a canonical scalar field $\phi$ with 
a slowly varying potential $V(\phi)$ (see \cite{review} for reviews).
For example, the simple power-law potential $V(\phi)=\lambda \phi^n/n$ 
($n$ and $\lambda$ are positive constants) leads to chaotic inflation for the field value
larger than the reduced Planck mass $M_{\rm pl}=2.435 \times 10^{18}$\,GeV \cite{Linde:1983gd}.
The quartic potential $V(\phi)=\lambda \phi^4/4$ is in tension 
with the WMAP constraints of the scalar spectral index $n_s$ and 
the tensor-to-scalar ratio $r$ \cite{Komatsu:2010fb}. 
Moreover the self coupling is constrained 
to be $\lambda \approx 10^{-13}$ from the WMAP 
normalization, which is much smaller than 
the typical coupling scale appearing 
in particle physics (e.g., $\lambda \approx 0.1$ 
for the Higgs boson \cite{Amsler:2008zzb}).

There are several different ways to reconcile the quartic
potential $V(\phi)=\lambda \phi^4/4$ with 
observations\footnote{In addition to a number of scenarios mentioned in Introduction, 
there is another way of realizing large $\lambda$ by using non-standard
kinetic terms \cite{Nakayama,DeTavakol,Varun}.}.
One of them is to introduce a non-minimal field coupling 
$\xi R \phi^2/2$ to the Ricci scalar $R$ \cite{Maeda,Bezrukov:2007ep}.
In the limit $\xi \gg 1$ the tensor-to-scalar ratio can be 
as small as $r \approx 10^{-3}$ with $n_s \approx 0.96$ \cite{Salopek}, 
which is well inside the $1\sigma$ observational contour \cite{Gumjudpai}. 
Moreover the self coupling is of the order of 
$\lambda \approx 10^{-10} \xi^2$ for $\xi \gg 1$ from the WMAP normalization.
If the field $\phi$ is a Higgs boson, however, this model is plagued by 
the problem of unitary violation around the energy scale of inflation \cite{unitarity}.
Moreover the non-minimal coupling $\xi R \phi^2/2$ does not necessarily
help other inflaton potentials to be compatible 
with observations \cite{Gumjudpai,DeTavakol}.

The second way is to use a non-minimal field derivative coupling to gravity
in the form $G^{\mu\nu}\partial_{\mu}\phi\partial_{\nu}\phi/(2M^2)$ 
\cite{Germani:2010gm}, 
where $G^{\mu\nu}$ is the Einstein tensor and $M$ is a mass scale
(see also Ref.~\cite{Amendola:1993uh} for the original work).
In the regime where the Hubble parameter $H$
is larger than $M$, the evolution of the field slows down 
due to a gravitationally enhanced friction.
In this case the potential $V(\phi)=\lambda \phi^4/4$ is compatible with the 
WMAP constraints of $n_s$ and $r$ with $\lambda \simeq 5.9 \times 10^{-32} 
(M_{\rm pl}/M)^4$ \cite{Tsujikawa:2012mk}.
Moreover the mechanism of slowing down the field 
(``slotheon'' \cite{slotheon}) works for general steep potentials.
For example this mechanism was applied to 
the potential $V(\phi)=\Lambda^4[1+\cos(\phi/f)]$ of natural inflation, 
where $\Lambda$ and $f$ are mass parameters \cite{GermaniNa}.
In conventional natural inflation \cite{Freese} the symmetry breaking scale $f$ 
needs to be larger than $3.5M_{\rm pl}$ for the consistency with 
the WMAP constraints \cite{Savage:2006tr}, but in this regime
standard quantum field theory is unlikely to be trustable \cite{Banks:2003sx}.
In the presence of the field derivative coupling, natural inflation can be 
compatible with the WMAP bounds even for $f$ smaller 
than $M_{\rm pl}$ \cite{Watanabe,Tsujikawa:2012mk}.

For the potential-driven inflation the field self-interaction of the form 
$(\partial\phi)^2\Box \phi$ \cite{Nicolis:2008in,Deffayet:2009wt,deRham}
also leads to the slow evolution of inflaton along the potential \cite{Kamada:2010qe}
(see Refs.~\cite{DPSV,G-de,G-inf} for the kinetically driven case).
The field equations of motion following from the Lagrangian 
$(\partial\phi)^2\Box \phi$ respects the the Galilean symmetry 
$\partial_{\mu} \phi \rightarrow \partial_{\mu} \phi+b_{\mu}$
in the limit of Minkowski space-time \cite{Nicolis:2008in}.
In a manifold having integrable (covariantly constant) Killing vectors $\xi^a$,
such a ``Galileon'' Lagrangian is invariant under the curved-space Galilean
transformation $\phi (x) \to \phi(x)+c+c_a \int_{x_0}^x \xi^a$, where 
$c$, $c_a$, $x_0$ are constants and $x$ is a space-time 
coordinate \cite{slotheon} (whose property also holds for the 
derivative coupling $G^{\mu\nu}\partial_{\mu}\phi\partial_{\nu}\phi$).
The presence of such a symmetry has an advantage that the theory 
can be quantum mechanically under control \cite{quantum}.

For the potential $V(\phi)=\lambda \phi^4/4$ the Galileon term 
$(\partial\phi)^2\Box \phi$ not only leads to the suppression 
of the tensor-to-scalar ratio compatible with recent observations, 
but also it gives rise to the coupling $\lambda$ of the order 
of 0.1 consistent with the WMAP 
normalization \cite{Kamada:2010qe,Popa,DeTavakol}. 
Meanwhile it is not clear whether the presence of such a non-linear 
field self-interaction does not disturb the oscillation of inflaton
during reheating. The absence of oscillations means that the 
standard mechanism of reheating (decays of inflaton to other 
particles and the thermalization of the Universe) does not work.
Moreover we need to check whether the conditions 
for the avoidance of ghosts and Laplacian instabilities can be avoided
after inflation. Since such conditions were recently derived 
in Refs.~\cite{Kobayashi:2011nu,Gao,DeFelice11,DeFelice:2011bh} 
for the most general scalar-tensor theories
having second-order equations of motion \cite{Horndeski,Deffayet11,Char}, 
those results can be applied to potential-driven inflation 
with the Galileon Lagrangian.

In this paper we study the dynamics of inflation and the subsequent reheating 
for the potentials $V(\phi)=\lambda \phi^n/n$ and 
$V(\phi)=\Lambda^4[1+\cos(\phi/f)]$ in the presence of the 
Galileon Lagrangian $(\partial\phi)^2\Box \phi$.
If the Galileon self-interaction dominates over the 
standard kinetic term after inflation, the oscillatory regime
of inflaton tends to disappear for both potentials. 
This is usually accompanied by a negative propagation 
speed squared $c_s^2$ of the scalar mode, which leads to the
instability of scalar perturbations on smaller scales. 
The model parameters of the potentials can be
constrained to have the coherent oscillation of inflaton
as well as to match with the observational data. 
For the quartic potential $V(\phi)=\lambda \phi^4/4$, 
for example, the self coupling $\lambda$ is bounded to be 
very much smaller than 1.
In natural inflation we show that it is difficult to realize 
the regime where the symmetry breaking scale $f$ is 
smaller than $M_{\rm pl}$.
We also study the effect of other covariant Galileon 
terms \cite{Deffayet:2009wt} on the dynamics of inflation and reheating
for the potentials $V(\phi)=\lambda \phi^n/n$.
It is possible to find some cases in which the self coupling of the 
potential $V(\phi)=\lambda \phi^4/4$ is of the order of 0.1, but 
the violent instability associated with negative $c_s^2$ is 
usually unavoidable.

This paper is organized as follows.
In Sec.~\ref{general} we present the background and perturbation equations for
potential-driven inflation in the presence of (generalized) Galileon Lagrangians.
The spectra of scalar and tensor perturbations are given 
by using slow-roll parameters.
In Sec.~\ref{G3} we study the models of chaotic inflation as well as natural inflation 
in the presence of the term $(\partial\phi)^2\Box \phi$ alone.
We clarify the viable parameter space in which the coherent oscillation 
of inflaton occurs during reheating. 
We also place observational constraints on the inflaton potentials
from the information of the scalar spectral index $n_s$ and 
the tensor-to-scalar ratio $r$.
In Secs.~\ref{G4} and \ref{G5} we provide similar constraints on the 
parameter space of chaotic inflation in the presence of other covariant 
Galileon terms.
Sec.~\ref{Conclusions} is devoted to conclusions.

\section{General field equations for the background and perturbations}
\label{general}

We start with the following action
\begin{equation}
S=\int d^4x \sqrt{-g} \left[ \frac{M_{\rm pl}^2}{2} R 
+ P(\phi,X)-G_3(\phi,X)\Box\phi+{\cal L}_4+{\cal L}_5 \right] \,,
\label{kinfaction}
\end{equation}
where $g$ is a determinant of the metric $g_{\mu \nu}$,
$R$ is a scalar curvature, and
\begin{eqnarray}
& & {\cal L}_4=G_4(\phi,X)R+G_{4,X}\left[ (\Box\phi)^2
-(\nabla_{\mu}\nabla_{\nu}\phi)(\nabla^{\mu}\nabla^{\nu}\phi) \right]\,, \\
& & {\cal L}_5=G_5(\phi,X)G_{\mu\nu}(\nabla^{\mu}\nabla^{\nu}\phi)-\frac{1}{6}G_{5,X}\left[ (\Box\phi)^3-3(\Box\phi)(\nabla_{\mu}\nabla_{\nu}\phi)(\nabla^{\mu}\nabla^{\nu}\phi)+2(\nabla^{\mu}\nabla_{\alpha}\phi)(\nabla^{\alpha}\nabla_{\beta}\phi)(\nabla^{\beta}\nabla_{\mu}\phi) \right]\,.
\end{eqnarray}
Here $P$ and $G_i \, (i=3,4,5)$ are functions in terms of $\phi$ and 
$X=-g^{\mu\nu} \partial_{\mu}\phi\partial_{\nu}\phi/2$
with the partial derivatives $G_{i,X}\equiv \partial G_i/\partial X$, and $G_{\mu\nu}=R_{\mu\nu}-g_{\mu\nu}R/2$ is the Einstein tensor 
($R_{\mu\nu}$ is the Ricci tensor). 
The action (\ref{kinfaction}) corresponds to the most general scalar-tensor 
theories having second-order equations of 
motion\footnote{Note that the cosmological dynamics in the presence of 
the general Lagrangian $G_3(\phi,X)\Box\phi$ was studied in Ref.~\cite{DPSV}
in the context of dark energy. In Refs.~\cite{G-de} the authors chose some 
particular forms of the function $G_3(\phi,X)$ to discuss the dynamics of
dark energy.} \cite{Deffayet11,Char,Kobayashi:2011nu}. 
This was first discovered by Horndeski in a different form \cite{Horndeski}.

We focus on the models in which inflation is mainly driven 
by a field potential $V(\phi)$, i.e., 
\begin{equation}
P(\phi,X)=X-V(\phi)\,.
\label{Plag}
\end{equation}
For the functions $G_i$ ($i=3,4,5$) we take
\begin{equation}
G_3(\phi,X)=f_3(\phi)X\,, \qquad
G_4(\phi,X)=f_4(\phi)X^2\,, \qquad
G_5(\phi,X)=f_5(\phi)X^2\,,
\label{Gi}
\end{equation}
where $f_i (\phi)$ depend on $\phi$ alone.
The covariant Galileon \cite{Deffayet:2009wt} corresponds 
to the choice \cite{DKT11}
\begin{equation}
f_3=\frac{c_3}{M^3}\,,\qquad
f_4=-\frac{c_4}{M^6}\,,\qquad
f_5=\frac{3c_5}{M^9}\,,
\label{fi}
\end{equation}
where $c_3$, $c_4$, $c_5$ are dimensionless constants, 
and $M$ is a constant having a dimension of mass.
We derive the background and perturbation equations
for the general functions (\ref{Gi}) in order to cover both 
the covariant Galileon and the coupling of the form
$f_3(\phi) \propto \phi$ discussed in Refs.~\cite{Kamada:2010qe,Kamada2}.
In principle we can extend the functions (\ref{Gi}) to more general forms (like 
the Horndeski's action \cite{Horndeski}), but our interest in this paper is to understand 
the effect of the Galileon-like self-interactions on the dynamics of  
inflation and reheating.
After Sec.~\ref{G3} we mainly focus on the covariant Galileon.

\subsection{Background equations}
\label{background}

On the flat Friedmann-Lema\^{i}tre-Robertson-Walker (FLRW) space-time
with the scale factor $a(t)$ (where $t$ is cosmic time) 
the background equations for the theories described by the functions 
(\ref{Plag}) and (\ref{Gi}) are given by \cite{Kobayashi:2011nu,DeFelice11}
\begin{eqnarray}
\hspace{-0.5cm} E_1 &\equiv& 3M_{\rm pl}^2 H^2-X-V-6H f_3\dot{\phi}X
+2\left( f_{3,\phi}-45H^2f_4 \right)X^2
+2H\left( 15f_{4,\phi}-14H^2f_5 \right)\dot{\phi}X^2+42H^2f_{5,\phi}X^3=0 \,, \label{E1} \\
\hspace{-0.5cm} E_2 &\equiv& 3M_{\rm pl}^2 H^2+X-V+2\left( M_{\rm pl}^2
-6f_4X^2-4Hf_5\dot{\phi}X^2+2f_{5,\phi}X^3 \right)\dot{H} \nonumber \\
&&-2\left[ f_3+12Hf_4\dot{\phi}-\left(5f_{4,\phi}-10H^2f_5\right)X
-6Hf_{5,\phi}\dot{\phi}X \right]X\ddot{\phi} \nonumber \\ 
&&-2\left( f_{3,\phi}+9H^2f_4 \right)X^2-4\left( 3f_{4,\phi}+2H^2f_5 \right)
H\dot{\phi}X^2+2\left( 2f_{4,\phi\phi}-H^2f_{5,\phi} \right)X^3+4Hf_{5,\phi\phi}\dot{\phi}X^3=0 \,,  \label{E2} \\
\hspace{-0.5cm} E_3 &\equiv& 
3H \dot{\phi}+V_{,\phi} +18H^2f_3X+108H^3f_4\dot{\phi}X
-2\left( f_{3,\phi\phi}+18H^2f_{4,\phi}-30H^4f_5 \right)X^2 \nonumber \\
& &+\left[ 1+6Hf_3\dot{\phi}-4\left( f_{3,\phi}-27H^2f_4 \right)X
-20\left( 3f_{4, \phi}-2H^2f_5 \right)H\dot{\phi}X-90H^2f_{5,\phi}X^2 \right]\ddot{\phi}  \nonumber \\
& &+2\left[ 3f_3+36Hf_4\dot{\phi}-\left( 15f_{4,\phi}-30H^2f_5 \right)X-18Hf_{5,\phi}\dot{\phi}X \right]X\dot{H}
 \nonumber \\
&&-2\left( 12f_{4,\phi\phi}+19H^2f_{5,\phi} \right)H\dot{\phi}X^2-30H^2f_{5,\phi\phi}X^3=0 \,,
\label{E3}
\end{eqnarray}
where $H\equiv \dot{a}/a$ is the Hubble parameter, and 
a dot denotes a derivative with respect to $t$.

Let us consider the covariant Galileon Lagrangian where the functions
$f_i$ ($i=3,4,5$) are given by Eq.~(\ref{fi}).  
In this case it is convenient to introduce 
the following dimensionless quantities
\begin{equation}
x=\frac{\phi}{M_{\rm pl}}\,, \qquad  y=\frac{\dot{\phi}}{MM_{\rm pl}}\,, 
\qquad z=\frac{H}{M} \,,
\end{equation}
and 
\begin{equation}
\tau=Mt\,,\qquad U(x)=\frac{V}{M^2M_{\rm pl}^2}\,, \qquad 
U_{,\phi}(x)=\frac{V_{,\phi}}{M^2M_{\rm pl}}\,, 
\qquad \alpha=\frac{M_{\rm pl}}{M} \,.
\label{dimendef}
\end{equation}
The constraint equation (\ref{E1}) can be written as 
\begin{equation}
6z^2-y^2-2U(x)-6c_3\alpha y^3 z+45c_4 \alpha^2 y^4 z^2
-42 c_5 \alpha^3 y^5 z^3=0\,.
\label{contsrainteq}
\end{equation}
Combining Eqs.~(\ref{E1}) and (\ref{E2}) to eliminate $V$
and then using Eq.~(\ref{E3}) to solve for $\ddot{\phi}$ and 
$\dot{H}$, it follows that 
\begin{eqnarray}
\frac{dx}{d\tau} &= &y \,, \label{eom1}\\
\frac{dy}{d\tau} &=& [9c_3^2 \alpha^2 y^5 z+3c_3 \alpha y^2
(78 c_5 \alpha^3 y^5 z^3-63 c_4 \alpha^2 y^4 z^2+y^2-6z^2)
+810 c_4^2 \alpha^4 y^7 z^3-3c_4 \alpha^2 y^3
(603 c_5 \alpha^3 y^5 z^4 \nonumber \\
& &+15y^2 z+U_{,\phi} (x) y-36z^3)+945c_5^2 \alpha^6
y^9 z^5+3c_5 \alpha^3 y^4 z (21y^2 z-30z^3 +2U_{,\phi}(x)y)
-2U_{,\phi}(x)-6yz]/\Delta\,,
\label{eom2} \\
\frac{dz}{d\tau} &=& -[27c_3^2 \alpha^2 y^2 z^2+
c_3 \alpha (450 c_5 \alpha^3 y^4 z^4-432 c_4 \alpha^2
y^3 z^3+12yz+U_{,\phi}(x))+1620c_4^2 \alpha^4 y^4 z^4
-12c_4 \alpha^2 yz (9yz \nonumber \\
& &+270c_5 \alpha^3 y^4 z^4+U_{,\phi} (x))
+1+1575 c_5^2 \alpha^6 y^6 z^6+15c_5 \alpha^3 y^2 z^2
(8yz+U_{,\phi}(x))]y^2/\Delta\,,
\label{eom3} 
\end{eqnarray}
where
\begin{eqnarray}
\Delta &=&
2+3c_3 \alpha y (4z+c_3 \alpha y^3)-3 c_4 \alpha^2 y^2
(36z^2 -y^2+18c_3 \alpha y^3 z-90c_4 \alpha^2 y^4 z^2 )
\nonumber \\
& &+3c_5 \alpha^3 y^3 z (40z^2-2y^2+18c_3 \alpha y^3 z
-192c_4 \alpha^2 y^4 z^2+105c_5 \alpha^3 y^5 z^3)\,.
\label{deter}
\end{eqnarray}
Numerically it is usually more stable to solve Eqs.~(\ref{eom1}) and (\ref{eom2})
with the constraint equation (\ref{contsrainteq}) rather than solving 
Eqs.~(\ref{eom1})-(\ref{eom3}).

\subsection{The spectra of density perturbations}

The spectra of scalar and tensor perturbations generated in the theories 
given by the action (\ref{kinfaction}) were derived in 
Refs.~\cite{Kobayashi:2011nu,Gao,DeFelice11}.
Here, we briefly review their formulas in order to apply 
them to concrete inflaton potentials.

The perturbed line element about the flat FLRW background 
is given by \cite{Co_Per}
\begin{equation}
ds^2=-(1+2A)dt^2+2\partial_iBdtdx^i+a^2(t)\left[ (1+2{\cal R})
\delta_{ij}+h_{ij} \right]dx^idx^j 
\label{metric} \,,
\end{equation}
where $A$, $B$, ${\cal R}$ are scalar metric perturbations, 
and $h_{ij}$ are tensor perturbations which are transverse and traceless.
The inflaton field is decomposed into the background and inhomogeneous parts, 
as $\phi=\phi_0(t)+\delta\phi(t,{\bm x})$.
We choose the uniform-field gauge characterized by $\delta\phi=0$, 
which fixes the time-component of a gauge-transformation vector $\xi^{\mu}$.
The scalar perturbation $E$, which appears as the form $E_{,ij}$ in the 
last term of (\ref{metric}), is gauged away, so that the spatial part
of $\xi^{\mu}$ is fixed.
Vector perturbations decay during inflation, 
so that their contribution is negligibly small.

We expand the action (\ref{kinfaction}) up to second-order in perturbations
by using the Hamiltonian and momentum constraints.
For the theories given by Eqs.~(\ref{Plag}) and (\ref{Gi})
the second-order action for scalar perturbations 
reduces to \cite{Kobayashi:2011nu,DeFelice11} 
\begin{equation}
S_s^{(2)}=\int dtd^3x \, a^3Q_s
\left[ \dot{{\cal R}}^2-\frac{c_s^2}{a^2}(\partial{\cal R})^2 \right]\,,
\end{equation}
where
\begin{equation}
Q_s=\frac{w_1(4w_1w_3+9w_2^2)}{3w_2^2}\,,\qquad
c_s^2=\frac{3(2w_1^2w_2H-w_2^2w_4+4w_1\dot{w_1}w_2-2w_1^2\dot{w_2})}
{w_1(4w_1w_3+9w_2^2)} \,,
\label{Qscs}
\end{equation}
and
\begin{eqnarray}
& & w_1=M_{\rm pl}^2-2\left( 3f_4+2Hf_5\dot{\phi} \right)X^2+2f_{5,\phi}X^3 \,, \\
& & w_2=2M_{\rm pl}^2H-2f_3\dot{\phi}X-2\left( 30Hf_4-5f_{4,\phi}\dot{\phi}
+14H^2f_5\dot{\phi} \right)X^2+28Hf_{5,\phi}X^3 \,, \\
& & w_3=-9M_{\rm pl}^2H^2+3\left( 1+12Hf_3\dot{\phi} \right)X+6\left( 135H^2f_4-2f_{3,\phi}-45Hf_{4,\phi}\dot{\phi}+56H^3f_5\dot{\phi} \right)X^2-504H^2f_{5,\phi}X^3 \,, \\
& & w_4=M_{\rm pl}^2+2\left( f_4-2f_5\ddot{\phi} \right)X^2-2f_{5,\phi}X^3 \,.
\end{eqnarray}
The conditions for the avoidance of ghosts and Laplacian instabilities 
correspond to $Q_s>0$ and $c_s^2>0$, respectively.
The two-point correlation function of the curvature perturbation ${\cal R}$ 
can be derived by employing the standard method of quantizing 
the fields on a quasi de Sitter background \cite{Co_Per}.
Using the solution for ${\cal R}$ obtained under the slow-roll
approximation, the power spectrum of the curvature perturbation is
\begin{equation}
{\cal P}_{s}=\frac{H^2}{8\pi^2 Q_s c_s^3} \,,
\end{equation}
which is evaluated at $c_sk=aH$ 
(where $k$ is a comoving wavenumber).

We decompose the intrinsic tensor perturbation $h_{ij}$
into two independent polarization modes, as
$h_{ij}=h_{+} e_{ij}^{+}+h_{\times} e_{ij}^{\times}$.
Then the second-order action for tensor perturbations 
is given by 
\begin{equation}
S_t^{(2)}=\sum_p \int dt\, d^3x \, 
a^3 Q_t \left[ \dot{h}_p^2 - \frac{c_t^2}{a^2}(\partial h_p)^2 \right] \,,
\end{equation}
where $p=+\,,\times$, and
\begin{equation}
Q_t=\frac{w_1}{4}\,,\qquad
c_t^2 = \frac{w_4}{w_1}\,.
\label{Qtct}
\end{equation}
We require that $Q_t>0$ and $c_t^2>0$ to avoid  
ghosts and Laplacian instabilities.
The tensor power spectrum is 
\begin{equation}
{\cal P}_t = \frac{H^2}{2\pi^2Q_tc_t^3} \,,
\end{equation}
which is evaluated at $c_tk=aH$.

\subsection{Slow-roll analysis}

For the covariant Galileon theory (\ref{fi}) we employ the slow-roll 
approximation to estimate the physical quantities introduced
in previous subsections.
Eliminating the term $V$ from Eqs.~(\ref{E1}) and (\ref{E2}), 
we obtain the equation for $\epsilon \equiv -\dot{H}/H^2$
expressed in terms of the slow-roll parameters
\begin{equation}
\delta_X=\frac{X}{M_{\rm pl}^2 H^2}\,,\quad
\delta_3=\frac{c_3 \dot{\phi}X}{M_{\rm pl}^2 M^3 H}\,,\quad
\delta_4=-\frac{2c_4 X^2}{M_{\rm pl}^2M^6}\,,\quad
\delta_5=\frac{6c_5 H \dot{\phi}X^2}{M_{\rm pl}^2 M^9}\,,\quad
\delta_{\phi}=\frac{\ddot{\phi}}{H \dot{\phi}}\,,
\label{del345}
\end{equation}
which are much smaller than unity during inflation.
It then follows that 
\begin{equation}
\epsilon=\frac{\delta_X+3\delta_3+18\delta_4+5\delta_5
-\delta_{\phi} (\delta_3+12\delta_4+5\delta_5)}
{1-3\delta_4-2\delta_5}
\simeq \delta_X+3\delta_3+18\delta_4+5\delta_5\,.
\label{epdef}
\end{equation}
In the second approximate equality we neglected the terms 
at second-order in slow-roll.

Under the slow-roll approximation the field equations (\ref{E1}) 
and (\ref{E3}) reduce to
\begin{eqnarray}
& & 3M_{\rm pl}^2 H^2 \simeq V\,,\label{slowba1} \\
& & 3H \dot{\phi} (1+{\cal A})+V_{,\phi} \simeq 0\,,
\label{slowba2}
\end{eqnarray}
where 
\begin{eqnarray}
{\cal A}=3c_3 \frac{H \dot{\phi}}{M^3}
-18 c_4 \left( \frac{H \dot{\phi}}{M^3} \right)^2
+15 c_5 \left( \frac{H \dot{\phi}}{M^3} \right)^3
=\frac{3\delta_3+18\delta_4+5\delta_5}{\delta_X}\,.
\label{Adef}
\end{eqnarray}
Using Eqs.~(\ref{slowba1}) and (\ref{slowba2}), the 
parameter $\delta_X$ can be estimated as
\begin{equation}
\delta_X \simeq \frac{\epsilon_{\phi}}{(1+{\cal A})^2}\,,
\label{delXslow}
\end{equation}
where 
\begin{equation}
\epsilon_{\phi}=\frac{M_{\rm pl}^2}{2}
\left( \frac{V_{,\phi}}{V} \right)^2\,.
\label{epphi}
\end{equation}
{}From Eqs.~(\ref{epdef}) and (\ref{Adef}) it follows that 
\begin{equation}
\epsilon \simeq (1+{\cal A}) \delta_{X} \simeq
\frac{\epsilon_{\phi}}{1+{\cal A}}\,.
\label{epap}
\end{equation}
The conventional slow-roll inflation corresponds to 
the limit ${\cal A} \to 0$, in which case $\epsilon \simeq 
\epsilon_{\phi} \simeq \delta_X$.
In the regime where $|{\cal A}|$ is much larger than 1
the evolution of the field slows down relative to that in 
standard inflation.

We define the number of e-foldings from the time $t$ 
to the time $t_f$ at the end of inflation, as 
$N=\int_t^{t_f} H (\tilde{t})\,d \tilde{t}$.
{}From Eqs.~(\ref{slowba1}) and (\ref{slowba2}) we have
\begin{equation}
N \simeq \frac{1}{M_{\rm pl}^2} \int_{\phi_f}^{\phi}
\left( 1+{\cal A} \right) \frac{V}{V_{,\tilde{\phi}}} 
d \tilde{\phi}\,.
\label{Ndef}
\end{equation}
The field value $\phi_f$ at the end of inflation is known by solving 
$\epsilon(\phi_f)=1$, that is
\begin{equation}
\epsilon_{\phi} (\phi_f)=1+{\cal A} (\phi_f)\,.
\label{phif}
\end{equation}
Since the factor ${\cal A}$ in Eq.~(\ref{Adef}) involves the field velocity, we need
to express $\dot{\phi}$ in terms of $\phi$ according to 
Eq.~(\ref{slowba2}) for the evaluation of $\phi_f$ and $N$.

Under the slow-roll approximation the quantities 
$Q_s$ and $c_s^2$ read
\begin{eqnarray}
Q_s & \simeq& M_{\rm pl}^2 \left( \delta_{X}+
6\delta_{3}+54 \delta_{4}+20\delta_{5} \right)\,,\\
c_s^2 &\simeq & \frac{\delta_X+4\delta_3+26\delta_4+8\delta_5}
{\delta_X+6\delta_3+54\delta_4+20\delta_5}\,.
\label{cs2ap}
\end{eqnarray}
In the regime where the Galileon self-interactions dominate over the
standard kinetic term we have 
$\{ |\delta_3|, |\delta_4|, |\delta_5| \} \gg \delta_X$.
In order to avoid that $Q_s$ becomes negative we demand 
the following conditions
\begin{equation}
c_3 \dot{\phi}>0\,,\qquad c_4<0\,,\qquad
c_5 \dot{\phi}>0\,.
\label{cicon}
\end{equation}
If $\delta_X$ is much larger than $|\delta_3|, |\delta_4|$,
and $|\delta_5|$, then the scalar propagation speed squared
is close to $1$. If either of $\delta_i$ ($i=1,2,3$) is the dominant 
contribution in Eq.~(\ref{cs2ap}), we have 
\begin{eqnarray}
c_s^2 &\simeq& 2/3 \qquad (\delta_3~{\rm dominant})\,,\\
c_s^2 &\simeq& 13/27  \qquad  (\delta_4~{\rm dominant})\,,\\
c_s^2 &\simeq& 2/5  \qquad  (\delta_5~{\rm dominant})\,.
\end{eqnarray}
This shows that the Laplacian instability of scalar perturbations 
is absent during slow-roll inflation.

The quantities $Q_t$ and $c_t^2$ are 
approximately given by 
\begin{equation}
Q_t \simeq \frac{M_{\rm pl}^2}{4} (1-3\delta_{4}-2\delta_5)\,,
\qquad 
c_t^2 \simeq 1+4\delta_4+2\delta_5\,,
\end{equation}
which are both positive.
Since we require that $\delta_4>0$ and $\delta_5>0$ to 
avoid scalar ghosts [see Eqs.~(\ref{del345}) and (\ref{cicon})], 
the tensor propagation speed squared is slightly superluminal 
in the presence of the couplings $G_4$ and $G_5$.

Under the slow-roll approximation the power spectra of 
scalar and tensor perturbations are given, respectively, by 
\begin{eqnarray}
{\cal P}_s &\simeq& \frac{H^2}{8\pi^2 M_{\rm pl}^2}
\frac{1}{c_s \epsilon_s} \simeq
\frac{V}{24\pi^2 M_{\rm pl}^4}
\frac{(\delta_X+6\delta_3+54\delta_4+20\delta_5)^{1/2}}
{(\delta_X+4\delta_3+26\delta_4+8\delta_5)^{3/2}}\,,
\label{Psf} \\
{\cal P}_t &\simeq& \frac{2H^2}{\pi^2 M_{\rm pl}^2} \simeq
\frac{2V}{3\pi^2 M_{\rm pl}^4}\,,
\end{eqnarray}
where 
\begin{equation}
\epsilon_s=\frac{Q_s c_s^2}{M_{\rm pl}^2}
\simeq \delta_X+4\delta_3+26\delta_4+8\delta_5\,.
\end{equation}
The tensor-to-scalar ratio is 
\begin{equation}
r=\frac{{\cal P}_t}{{\cal P}_s}=16c_s \epsilon_s
=16 \frac{(\delta_X+4\delta_3+26\delta_4+8\delta_5)^{3/2}}
{(\delta_X+6\delta_3+54\delta_4+20\delta_5)^{1/2}}\,.
\label{rgene}
\end{equation}
Defining the spectral indices as $n_s-1=d\ln {\cal P}_s/d\ln k|_{c_sk=aH}$
and $n_t=d\ln {\cal P}_t/d\ln k|_{c_tk=aH}$, it follows that 
\begin{eqnarray}
n_s-1 &=& -2\epsilon-\eta_s-s\,,\\
n_t &=& -2\epsilon\,,
\end{eqnarray}
where $\epsilon$ is given in Eq.~(\ref{epdef}), and 
\begin{equation}
\eta_s=\frac{\dot{\epsilon_s}}{H \epsilon_s}\,,\qquad
s=\frac{\dot{c_s}}{H c_s}\,.
\end{equation}

The consistency relation between $r$ and $n_t$ is 
\begin{equation}
r=-8c_s (n_t-2\delta_3-16\delta_4-6\delta_5)\,.
\end{equation}
In the regime $\delta_X \gg |\delta_i|$ ($i=1,2,3$) 
the standard consistency relation $r=-8n_t$ holds.
If either of the terms $|\delta_i|$ ($i=1,2,3$) dominates
over other terms, it follows that 
\begin{eqnarray}
r &=& -8.71 n_t \qquad (\delta_3~{\rm dominant})\,,\\
r &=& -8.02 n_t  \qquad  (\delta_4~{\rm dominant})\,,\\
r &=& -8.10 n_t \qquad  (\delta_5~{\rm dominant})\,.
\end{eqnarray}
Since the ratio $r/n_t$ is close to $-8$ in all cases, 
the observational bounds on $n_s$ and $r$ are similar to 
those derived by using the consistency relation $r=-8n_t$.

\section{Theories with $G_3\neq 0, G_4=0, G_5=0$}
\label{G3}

We first study the covariant Galileon theory in which only 
the term $-(c_3/M^3)X \square \phi$ is present in the action 
(\ref{kinfaction}), i.e., 
\begin{equation}
c_3 \neq 0\,,\qquad c_4=0\,,\qquad c_5=0\,.
\end{equation}
Solving the slow-roll equation (\ref{slowba2}) for $\dot{\phi}$, 
it follows that 
\begin{equation}
\dot{\phi}=\frac{M^3}{6c_3 H} 
\left( \sqrt{1-\frac{4c_3V_{,\phi}}{M^3}}-1 \right)\,,\qquad
{\rm and} \qquad
{\cal A}(\phi)=\frac12 
\left( \sqrt{1-\frac{4c_3V_{,\phi}}{M^3}}-1 \right)\,.
\label{dotphiG3}
\end{equation}
For $c_3>0$ one has $\dot{\phi}>0$ and $V_{,\phi}<0$ from 
Eqs.~(\ref{cicon}) and (\ref{slowba2}).
If $c_3<0$, then $\dot{\phi}<0$ and $V_{,\phi}>0$.
In the former and latter cases we choose the coefficients
$c_3=1$ and $c_3=-1$, respectively, without loss of generality.
The transition from Galileon inflation to standard inflation 
can be quantified by the condition ${\cal A}(\phi_G)=1$, 
which translates into
\begin{equation}
c_3 V_{,\phi} (\phi_G)=-2M^3\,.
\label{phiG}
\end{equation}

The field value $\phi_f$ at the end of inflation is 
known from Eq.~(\ref{phif}), i.e., 
\begin{equation}
\epsilon_{\phi} (\phi_f)=\frac12 \left[ 1
+\sqrt{1-\frac{4c_3 V_{,\phi} (\phi_f)}{M^3}}
\right]\,.
\label{phif2}
\end{equation}
For the scalar potential with $V_{,\phi}>0$
the transition from the regime $\delta_3 \gg \delta_X$ to
the regime $\delta_3 \ll \delta_X$ 
occurs during inflation provided that $|\phi_G|>|\phi_f|$.
On the other hand, if $|\phi_G|<|\phi_f|$, the Galileon 
self-interaction dominates over the standard kinetic
term during the whole stage of inflation.

Since $c_3V_{,\phi}/M^3=-{\cal A}(1+{\cal A})$, 
the number of e-foldings (\ref{Ndef}) reads
\begin{equation}
N=  -\frac{c_3}{M^3 M_{\rm pl}^2} 
\int_{\phi_f}^{\phi} 
\frac{V (\tilde{\phi})}{{\cal A}(\tilde{\phi})}
d\tilde{\phi}\,.
\label{efoldG3}
\end{equation}
Using Eqs.~(\ref{delXslow}), (\ref{epphi}), and the relation 
$\delta_3=({\cal A}/3)\delta_X$,
the scalar power spectrum (\ref{Psf}) reduces to
\begin{equation}
{\cal P}_s 
=\frac{V^3}{12\pi^2 M_{\rm pl}^6 V_{,\phi}^2}
\frac{(1+{\cal A})^2 (1+2{\cal A})^{1/2}}
{(1+4{\cal A}/3)^{3/2}}\,.
\label{PsG3}
\end{equation}
For a given inflaton potential and a mass scale $M$, 
the field value $\phi_f$ is known by solving Eq.~(\ref{phif2}).
Then the number of e-foldings can be evaluated from 
Eq.~(\ref{efoldG3}) to find the value of $\phi$ at $N=60$
(for which we denote $\phi_{60}$). 
The WMAP normalization of the scalar power spectrum 
is ${\cal P}_s (\phi_{60})=2.4 \times 10^{-9}$ \cite{Komatsu:2010fb},
by which the mass $M$ can be related to model parameters 
of the potential.

The scalar spectral index $n_s$ is known from Eq.~(\ref{PsG3})
according to $n_s-1=\dot{{\cal P}_s}/(H{\cal P}_s)$.
Taking the time derivative of the quantity ${\cal A}$ given in 
Eq.~(\ref{dotphiG3}) and using Eq.~(\ref{slowba1}), it follows
that $\dot{{\cal A}}/H=-\eta_{\phi} {\cal A}/(1+2{\cal A})$, where
\begin{equation}
\eta_{\phi}=M_{\rm pl}^2 \frac{V_{,\phi \phi}}{V}\,.
\label{etaphi}
\end{equation}
Then we obtain 
\begin{equation}
n_s-1=-\frac{6\epsilon_{\phi}}{1+{\cal A}}+
\frac{2\eta_{\phi}}{1+4{\cal A}/3} \left[ 1-
\frac{{\cal A}}{6 (1+2{\cal A})^2} \right]\,,
\label{nsG3}
\end{equation}
where $\epsilon_{\phi}$ is defined in Eq.~(\ref{epphi}).
For ${\cal A} \to 0$ the formula (\ref{nsG3})
recovers the result $n_s-1 \simeq -6\epsilon_{\phi}+2 \eta_{\phi}$
in conventional slow-roll inflation.
In the limit ${\cal A} \gg 1$ we have that 
$n_s-1 \simeq -6\epsilon_{\phi}/{\cal A}+3\eta_{\phi}/(2{\cal A})$.
{}From Eq.~(\ref{rgene}) the tensor-to-scalar ratio reads
\begin{equation}
r=16\epsilon_{\phi}
\frac{(1+4{\cal A}/3)^{3/2}}{(1+{\cal A})^2 (1+2{\cal A})^{1/2}}\,.
\label{rG3}
\end{equation}
In the limit ${\cal A} \to 0$ this reproduces the standard relation 
$r \simeq 16\epsilon_{\phi}$, but for ${\cal A} \gg 1$ we have 
$r \simeq 64\sqrt{6}\,\epsilon_{\phi}/(9{\cal A})$.
The observables (\ref{nsG3}) and (\ref{rG3}) are functions of 
$\phi(N)$, so that they can be evaluated for 
a given inflaton potential.

If the Galileon term is dominant even after the end of inflation, 
this affects the oscillation of inflaton during reheating.
In order to avoid that the $1+6H f_3 \dot{\phi}$ term in front of
$\ddot{\phi}$ in Eq.~(\ref{E3}) become negative, we require
\begin{equation}
1+6H c_3 \dot{\phi}/M^3>0\,.
\label{concon}
\end{equation}
The field velocity $\dot{\phi}$ changes its sign during 
the oscillating stage of inflaton.
This means that the condition (\ref{concon}) can be violated 
depending on the model parameters.
Note that the determinant (\ref{deter}) is approximately 
given by $\Delta \simeq 2(1+6H c_3 \dot{\phi}/M^3)$, 
so that the violation of the condition (\ref{concon})
leads to the divergence of Eqs.~(\ref{eom2}) and 
(\ref{eom3})\footnote{Note that a similar determinant singularity 
appears in the context of anisotropic string cosmology \cite{Toporensky}.}.
In the regime $1+6H c_3 \dot{\phi}/M^3<0$ the field climbs
up the potential like a phantom field, so successful reheating 
cannot be realized.

For smaller values of $M$ the condition (\ref{concon})
tends to be violated. In this case we also show that $c_s^2$ 
can be negative due to the dominance of the Galileon term.
For the potentials (a) $V(\phi)=\lambda \phi^n/n$ (chaotic inflation) 
and (b) $V(\phi)=\Lambda^4 [1+\cos (\phi/f)]$ (natural inflation),
we clarify the parameter space in which inflaton oscillates coherently  
and $c_s^2$ remains positive. 
We also place observational bounds on each model
from the information of the scalar spectral index $n_s$ 
and the tensor-to-scalar ratio $r$.

\subsection{Chaotic inflation}

First we study the case of chaotic inflation characterized by the potential
\begin{equation}
V(\phi)=\frac{\lambda}{n}\phi^n\,,
\label{chaoticpo}
\end{equation}
where $n$ and $\lambda$ are positive constants.
The initial value of the field $\phi$ is assumed to be positive, so 
that $\dot{\phi}<0$ and hence $c_3=-1$.
{}From Eq.~(\ref{phiG}) the field value at the transition 
is given by 
\begin{equation}
\phi_G= \left( 2M^3/\lambda \right)^{1/(n-1)}\,.
\end{equation}
If the slow-roll parameter (\ref{epap}) at $\phi=\phi_G$
is smaller than 1, the transition from Galileon inflation to standard inflation 
occurs during inflation. 
The condition $\epsilon(\phi_G)<1$ translates into
\begin{equation}
M> 2^{-n/3} n^{(n-1)/3} M_{\rm pl}^{(n-1)/3} 
\lambda^{1/3} \equiv M_c\,.
\label{Mcon}
\end{equation}

For the potential (\ref{chaoticpo}) the function ${\cal A}(\phi)$
in Eq.~(\ref{dotphiG3}) is given by 
\begin{equation}
{\cal A} (\phi)=\frac12 \left( \sqrt{1+\frac{4 \lambda \phi^{n-1}}
{M^3}}-1 \right)\,,
\end{equation}
in which case, apart from the case $n=2$,
the number of e-foldings (\ref{efoldG3})
is not integrated analytically. Moreover, in order to find
the field value $\phi_f$, we need to solve 
Eq.~(\ref{phif2}) numerically.
In the limit ${\cal A} \gg 1$, however, it is possible 
to derive the analytic expression of $\phi$ in terms of $N$.
Since ${\cal A}(\phi) \simeq \sqrt{\lambda \phi^{n-1}/M^3}$
in this limit, we have $\phi_f^{(n+3)/2} \simeq n^2 M_{\rm pl}^2 M^{3/2}
/(2\sqrt{\lambda})$ from Eq.~(\ref{phif2}).
Then, integration of Eq.~(\ref{efoldG3}) gives 
\begin{equation}
\phi^{(n+3)/2} \simeq \frac{nM_{\rm pl}^2 M^{3/2}}{2\sqrt{\lambda}}
\left[ (n+3)N+n \right]\,.
\end{equation}
Substituting this solution into Eq.~(\ref{PsG3}) and using the WMAP
normalization ${\cal P}_s=2.4 \times 10^{-9}$ at $N=60$, we find 
\begin{equation}
\lambda\,\frac{M^n}{M_{\rm pl}^4}=8^{(n+1)/3}n^2
\frac{(3.1 \times 10^{-7})^{(n+3)/3}}
{(61n+180)^{n+1}}\,.
\label{WMAP1}
\end{equation}
For smaller $M$, $\lambda$ tends to be larger.
The scalar spectral index (\ref{nsG3}) and 
the tensor-to-scalar ratio (\ref{rG3}) 
reduce to 
\begin{equation}
n_s=1-\frac{3(n+1)}{(n+3)N+n} \,, \qquad 
r=\frac{64 \sqrt{6}}{9}\frac{n}{(n+3)N+n} \,, 
\label{nsrga}
\end{equation}
which agree with those given in Ref.~\cite{Kamada:2010qe}.
For $n=4$ and $N=60$, for example,  
we have $n_s=0.965$ and $r=0.164$.

In the limit ${\cal A} \ll 1$ we have $\phi_f=nM_{\rm pl}/\sqrt{2}$
and $\phi^2=2nM_{\rm pl}^2 (N+n/4)$. The WMAP normalization 
at $N=60$ gives 
\begin{equation}
\lambda=2.8 \times 10^{-7}n^3 \left[ n(120+n/2) \right]^{-(n+2)/2}
M_{\rm pl}^{4-n}\,.
\label{WMAP2}
\end{equation}
The scalar spectral index and the tensor-to-scalar ratio are
\begin{equation}
n_s=1-\frac{2(n+2)}{4N+n} \,, \qquad 
r=\frac{16n}{4N+n} \,,
\label{nsrga2}
\end{equation}
which correspond to those for standard chaotic inflation.
In the regime between ${\cal A} \gg 1$ and ${\cal A} \ll 1$
we need to evaluate $n_s$ and $r$ numerically.
For $n=4$ and $N=60$, for example,  
we have $n_s=0.951$ and $r=0.262$.

\subsubsection{$V(\phi)=m^2 \phi^2/2$}
%

\begin{figure}
\includegraphics[height=3.3in,width=3.3in]{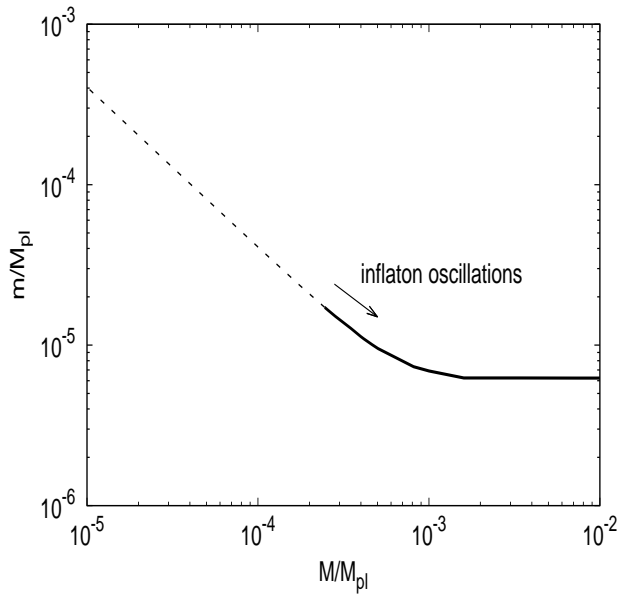}
\caption{\label{fig1}
The mass parameters $M$ and $m$ satisfying the WMAP normalization 
${\cal P}_s=2.4 \times 10^{-9}$ at $N=60$ for the quadratic potential 
$V(\phi)=m^2 \phi^2/2$ with the term $G_3=-X/M^3$.
The solid line represents the region in which the coherent 
oscillation occurs during reheating.}
\end{figure}

\begin{figure}
\includegraphics[height=3.1in,width=3.3in]{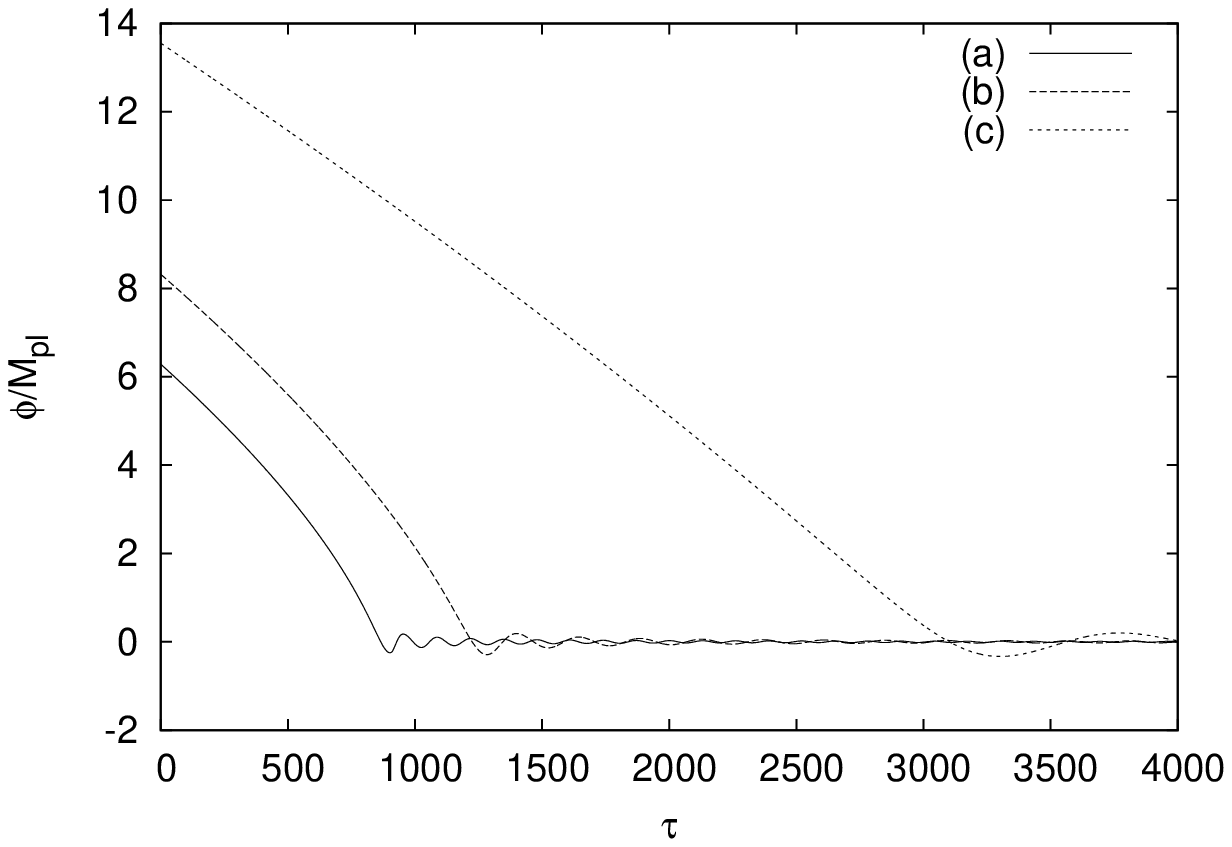}
\includegraphics[height=3.1in,width=3.3in]{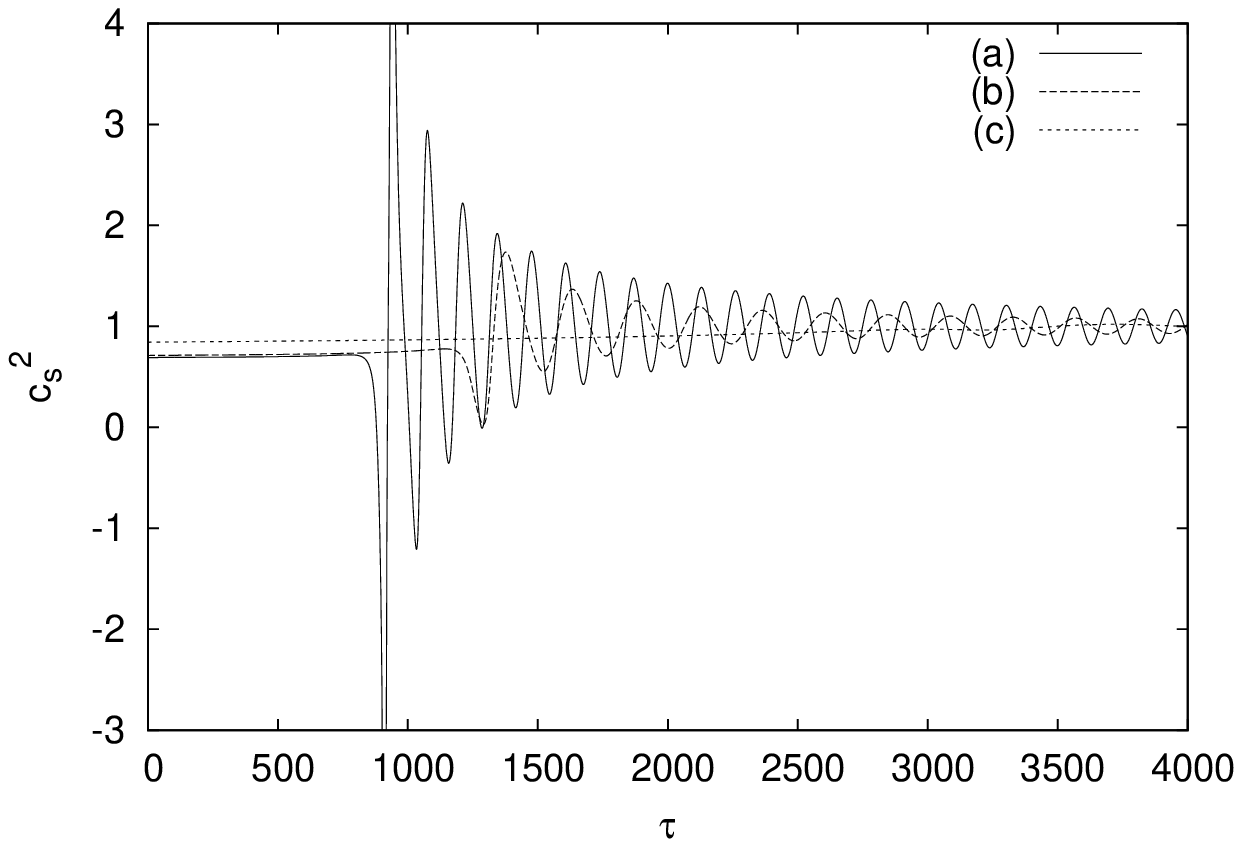}
\caption{\label{fig2}
Evolution of the field $\phi$ (left) and the scalar propagation 
speed squared $c_s^2$ (right) for the quadratic potential 
$V(\phi)=m^2 \phi^2/2$ with the term $G_3=-X/M^3$ in three different cases:
(a) $M=3.0 \times 10^{-4}M_{\rm pl}$, $m=1.45 \times 10^{-5}M_{\rm pl}$,
(b) $M=4.2 \times 10^{-4}M_{\rm pl}$, $m=1.1 \times 10^{-5}M_{\rm pl}$, 
and 
(c) $M=1.0 \times 10^{-3}M_{\rm pl}$, $m=6.9 \times 10^{-6}M_{\rm pl}$.
We choose the initial conditions at $N=60$ determined 
by the slow-roll analysis, i.e.,
(a) $x_i=6.28$, $y_i=-5.25 \times 10^{-3}$, $z_i=1.24 \times 10^{-1}$, 
(b) $x_i=8.32$, $y_i=-5.07 \times 10^{-3}$, $z_i=8.89 \times 10^{-2}$, 
(c) $x_i=13.55$, $y_i=-3.90 \times 10^{-3}$, $z_i=3.82 \times 10^{-2}$, 
respectively. 
In the case (a) the system enters the region with negative values of 
$c_s^2$, whereas in the case (c) $c_s^2$ is always positive.
The case (b) is the marginal one in which the minimum value 
of $c_s^2$ is 0.
}
\end{figure}

\begin{figure}
\includegraphics[height=3.5in,width=3.3in]{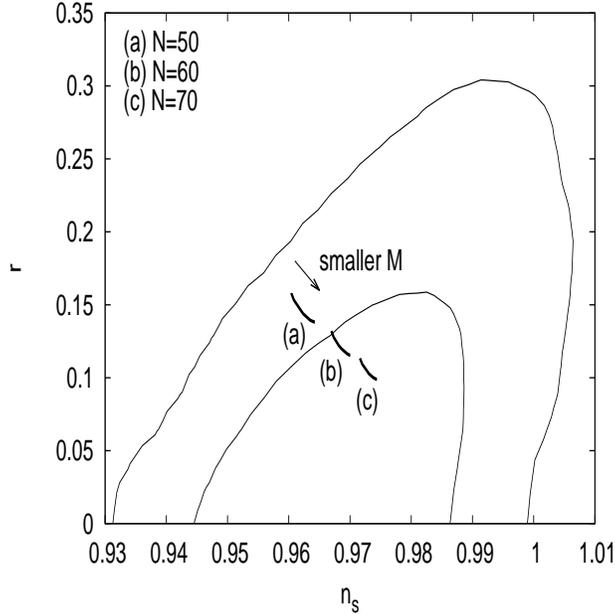}
\caption{\label{fig3}
Observational constraints on the quadratic potential 
$V(\phi)=m^2\phi^2/2$ with the term $G_3=-X/M^3$
in the $(n_s\,, r)$ plane for the 
numbers of e-foldings $N=50\,, 60\,, 70$.
The thin solid curves show the $1\sigma$ (inside) and 
$2\sigma$ (outside) observational contours constrained 
by the joint data analysis of WMAP7, BAO, and HST.
For smaller values of $M$ the tensor-to-scalar ratio $r$ gets smaller, 
whereas the scalar spectral index $n_s$ increases.
}
\end{figure}

Let us consider the case of the quadratic potential $V(\phi)=m^2\phi^2/2$, 
i.e., $\lambda=m^2$ and $n=2$.
In the limit ${\cal A} \gg 1$ the WMAP normalization (\ref{WMAP1})
leads to the following relation 
\begin{equation}
\frac{m}{M_{\rm pl}} \frac{M}{M_{\rm pl}}  \simeq 4.1 \times 10^{-9}\,.
\label{mM}
\end{equation}
In another limit ${\cal A} \ll 1$, we have 
$m \simeq 6.2 \times 10^{-6}M_{\rm pl}$ from Eq.~(\ref{WMAP2}).
In Fig.~\ref{fig1} we plot $m$ versus $M$ constrained by
the normalization ${\cal P}_s=2.4 \times 10^{-9}$
at $N=60$. For small $M$ satisfying $M/M_{\rm pl} \ll 10^{-3}$
the numerical result in Fig.~\ref{fig1} is in good agreement with 
the analytic estimation (\ref{mM}).
In the regime $M/M_{\rm pl} \gg 10^{-3}$ the mass $m$ approaches
the constant value $m \simeq 6.2 \times 10^{-6}M_{\rm pl}$.
Under the condition (\ref{Mcon}), i.e., $M>2^{-1/3}M_{\rm pl}^{1/3}m^{2/3}$, 
the transition from the regime ${\cal A}>1$ to the regime ${\cal A}<1$
occurs during inflation.
Combining this condition with the constraints on $M$ and $m$ 
shown in Fig.~\ref{fig1}, it follows that $M>4.0 \times 10^{-4}M_{\rm pl}$.
If $M<4.0 \times 10^{-4}M_{\rm pl}$, the Galileon self-interaction 
dominates over the standard kinetic term during inflation.

In order to see the effect of the Galileon term during inflation and reheating, 
we numerically solve the background equations (\ref{contsrainteq})-(\ref{eom2})
with the initial conditions determined by the slow-roll analysis.
We confirm that the slow-roll approximation is accurate enough to reproduce
the numerical values of $N$ with the difference less than a few percent.

In Fig.~\ref{fig2} we plot the evolution of $\phi$ and $c_s^2$ for 
three different mass parameters $M$ and $m$ constrained by the 
WMAP normalization. 
The case (a) corresponds to the mass $M=3.0 \times 10^{-4}M_{\rm pl}$, 
which is smaller than the critical mass $M_c=4.0 \times 10^{-4}M_{\rm pl}$.
Hence the Galileon self-interaction dominates over the standard kinetic term
by the end of inflation.
As we see in the right panel of Fig.~\ref{fig2}, the solutions enter the regime 
in which $c_s^2$ is negative. 
For $M$ smaller than $3.0 \times 10^{-4}M_{\rm pl}$
the period in which $c_s^2$ is negative tends to be 
longer with $|c_s^2|$ much larger than 1. 
Since scalar perturbations grow very rapidly in such cases, 
the Universe becomes inhomogeneous at the level of destroying  
the homogenous background.
The case (b) shown in Fig.~\ref{fig2} corresponds to the marginal one 
in which the minimum value of $c_s^2$ is 0.
In the case (c) the transition from the regime ${\cal A}>1$ 
to the regime ${\cal A}<1$ occurs during inflation and $c_s^2$
always remains positive.
For the range of masses $M$ used in the numerical simulations 
of Fig.~\ref{fig2}, the inflaton oscillates coherently 
as long as the backreaction of
created particles is neglected.

The condition for the avoidance of negative values of $c_s^2$ is
\begin{equation}
M>4.2 \times 10^{-4}M_{\rm pl}\,,
\label{con1}
\end{equation}
under which $c_s^2$ finally approaches 1 
without entering the regime $c_s^2<0$.
Note that $c_t^2=1$ in the presence of the $G_3$ term alone.
Numerically we find that inflaton oscillates
coherently during reheating for 
\begin{equation}
M>2.5 \times 10^{-4}M_{\rm pl}\,,
\label{con2}
\end{equation}
which is related to the condition (\ref{concon}).
During inflation in which $c_3 \dot{\phi}$ is always 
positive, the condition (\ref{concon}) is always satisfied.
However, after $\dot{\phi}$ changes its sign during 
reheating, the condition (\ref{concon}) is violated 
for $M<2.5 \times 10^{-4}M_{\rm pl}$.
The criterions (\ref{con1}) and (\ref{con2}) are not 
very different from each other.
We also confirmed that the conditions $Q_s>0$ and 
$Q_t>0$ are satisfied in such cases.

The superluminal behavior of the scalar propagation 
speed seen in Fig.~\ref{fig2} is a matter of 
debate \cite{super1,super2,super2d,super3,super4}.
This behavior does not necessarily imply a violation 
of causality because general solutions of Galileon 
models break Lorentz symmetry on the FLRW background. 
A problem occurs if closed time-like curves (CTCs) 
are developed by the existence of such a superluminal mode.
Hawking argued that the formation of CTCs may be generally 
avoided because the backreaction from the energy-momentum tensor 
of a quantum field becomes so large before the onset of 
formation of the CTC (which is called chronology 
protection conjecture) \cite{Hawking}.
According to the acoustic analogue of the chronology protection 
conjecture, Refs.~\cite{super2} showed that CTCs do not form 
even in the presence of the superluminal mode in k-essence
theories. In Ref.~\cite{super3} it was claimed that 
in Galileon theories the CTCs appear only when there exists some region 
in which higher derivative Galileon terms are larger than the 2-derivative 
kinetic term. 
On the other hand, Ref.~\cite{super4} showed that 
the CTCs do not arise because the Galileons become strongly coupled 
at the onset of formation of a CTC.
In our work we do not put the bounds $c_s^2 \le 1$
and $c_t^2 \le 1$ by taking an attitude that the existence of 
superluminal modes does not pose a problem
associated with the CTCs.

In Fig.~\ref{fig3} the theoretical values of $n_s$ and $r$ 
are plotted as a function of $M$ ranging in the region (\ref{con2})
with three different values of $N$ ($=50, 60, 70$).
We also show the $1\sigma$ and $2\sigma$ observational contours 
constrained by the joint data analysis of WMAP7 \cite{Komatsu:2010fb}, 
Baryon Acoustic Oscillations (BAO) \cite{Percival:2009xn}, and 
the Hubble constant measurement using the the Hubble Space 
Telescope (HST) \cite{Riess:2009pu}.
As we decrease the value of $M$, the two observables shift 
from the values in Eq.~(\ref{nsrga2}) to those in Eq.~(\ref{nsrga}).
For smaller $M$, $r$ gets smaller whereas $n_s$ increases, 
so that the quadratic potential shows better compatibility with the data.
Even for the mass $M$ corresponding to the lower limit of Eq.~(\ref{con2}) 
the term ${\cal A}$ is larger than 1 during most stage of inflation, 
in which case $n_s$ and $r$ are close to the asymptotic values given in 
Eq.~(\ref{nsrga}).

\subsubsection{$V(\phi)=\lambda \phi^4/4$}

\begin{figure}
\includegraphics[height=3.2in,width=3.3in]{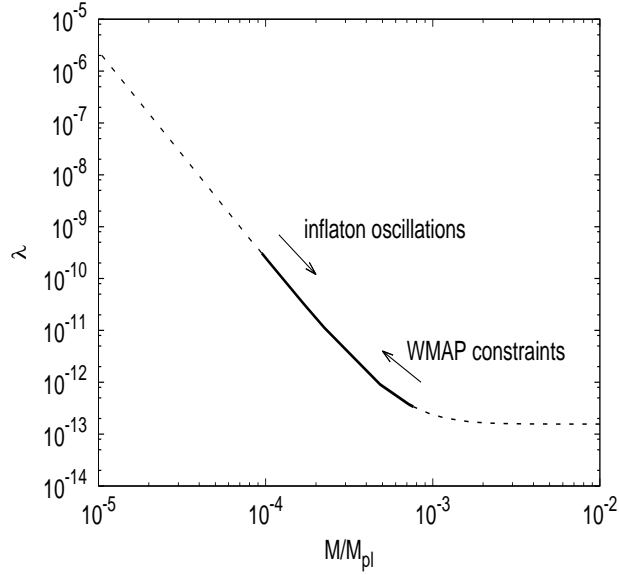}
\caption{\label{fig4}
The parameters $M$ and $\lambda$ satisfying the WMAP normalization 
${\cal P}_s=2.4 \times 10^{-9}$ at $N=60$ for the quartic potential 
$V(\phi)=\lambda \phi^4/4$ with the term $G_3=-X/M^3$.
The solid line represents the region in which 
the inflaton oscillation occurs during reheating
and the model is within the $2\sigma$ observational
contour in the ($n_s, r$) plane.}
\end{figure}

We proceed to the case of the quartic potential $V(\phi)=\lambda \phi^4/4$.
In the limits ${\cal A} \gg 1$ and ${\cal A} \ll 1$ the WMAP normalizations 
(\ref{WMAP1}) and (\ref{WMAP2}) give 
$\lambda (M/M_{\rm pl})^4 \simeq 2.4 \times 10^{-26}$ 
and $\lambda \simeq 1.6 \times 10^{-13}$, respectively.
Figure 4 shows the viable parameter space in the 
$(M, \lambda)$ plane satisfying the WMAP normalization at $N=60$.
Under the condition (\ref{Mcon}), i.e., $M>2^{2/3}\lambda^{1/3}M_{\rm pl}$, 
the transition from the regime ${\cal A}>1$ to the regime ${\cal A}<1$
occurs during inflation.
Combining this condition with the constraints on $M$ and $\lambda$ 
shown in Fig.~\ref{fig4}, it follows that $M>2.8 \times 10^{-4}M_{\rm pl}$.
If $M<2.8 \times 10^{-4}M_{\rm pl}$, the Galileon term dominates 
over the standard kinetic term during inflation.

By solving the background equations of motion 
(\ref{contsrainteq})-(\ref{eom2}), we find that $c_s^2$ 
remains positive for
\begin{equation}
M>1.7\times10^{-4}M_{\rm pl}\,.
\label{con1self}
\end{equation}
The inflaton oscillation occurs during reheating provided that   
\begin{equation}
M>9.5\times10^{-5}M_{\rm pl}\,.
\label{con2self}
\end{equation}
%

\begin{figure}
\includegraphics[height=3.3in,width=3.2in]{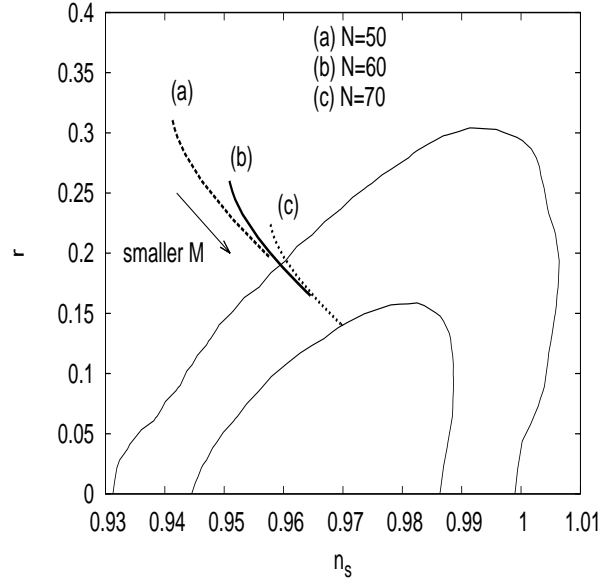}
\caption{\label{fig5}
Observational constraints on the quartic potential 
$V(\phi)=\lambda\phi^4/4$ with the term $G_3=-X/M^3$ 
for three different values of $N$.
The $1\sigma$ and $2\sigma$ observational contours 
are the same as those in Fig.~\ref{fig3}. 
While the standard case ($M \to \infty$) is outside the $2\sigma$ bound, 
the presence of the Galileon term 
can make the quartic potential compatible with observations.
}
\end{figure}

In Fig.~\ref{fig5} the theoretical values of $n_s$ and $r$ are plotted 
as a function of $M$ ranging in the region (\ref{con2self})
with $N = 50,60,70$. 
In the limit $M \to \infty$ the quartic potential is outside the 
$2\sigma$ observational contour for $N$ smaller than $70$.
In the presence of the Galileon term the model can be 
compatible with the current observations due to the suppressed 
tensor-to-scalar ratio and the larger scalar spectral index. 
For $N = 60$ the model is within the $2\sigma$ contour for  
\begin{equation}
M <7.7\times10^{-4}M_{\rm pl}\,.
\label{con3self}
\end{equation}
In terms of the parameter $\lambda$ the conditions (\ref{con2self}) 
and (\ref{con3self}) translate into
\begin{equation}
3.4\times10^{-13}<\lambda <3.1\times10^{-10} \,.
\label{lam_G3}
\end{equation}
Under the constraint (\ref{con1self}) the upper bound
is $\lambda<3.0 \times 10^{-11}$.
The result (\ref{lam_G3}) shows that one cannot accommodate
the self coupling $\lambda \sim 0.1$ of the Higgs boson
in the presence of the coupling $G_3=-X/M^3$.

We also studied the case of the generalized Galileon 
term $-G_3(\phi, X) \square \phi$, where
\begin{equation}
G_3=\frac{c_3}{M^4} \phi X\,,
\end{equation}
which was proposed in Ref.~\cite{Kamada:2010qe}.
Numerically we find that the inflaton oscillation 
occurs for $M>3.6\times10^{-4}M_{\rm pl}$ 
and $\lambda<2.7\times10^{-8}$.
The self coupling $\lambda$ is still much smaller 
than the order of 0.1.

\subsection{Natural inflation}

Natural inflation \cite{Freese} is characterized 
by the potential
\begin{equation}
V(\phi)=\Lambda^4\left[ 1+\cos\left( \frac{\phi}{f} \right) \right] \,, 
\label{natural_pot}
\end{equation}
where $\Lambda$ and $f$ are constants having the dimension of mass. 
In the absence of the Galileon term this potential can be 
compatible with the observational data only for 
$f \gtrsim 3.5M_{\rm pl}$ \cite{Savage:2006tr}.
This is the regime in which standard quantum field theory 
may not be reliable. If the field $\phi$ is a string axion, 
$f$ is usually smaller than the order of 
$M_{\rm pl}$ \cite{Banks:2003sx,multi_natudal}.
In the following we study whether this problem can be alleviated or not 
in the presence of the Galileon term $G_3=c_3X/M^3$.

We assume that inflation occurs in the region $0<\phi/f<\pi$, 
in which case $\dot{\phi}>0$. 
We choose $c_3=1$ to satisfy the condition (\ref{cicon}).
For the potential (\ref{natural_pot}) the function ${\cal A}(\phi)$
in Eq.~(\ref{dotphiG3}) is given by 
\begin{equation}
{\cal A} (\phi)=\frac12 \left( \sqrt{1+
\frac{4 \gamma \sin(\phi/f)}{q}} -1 \right)
\label{A_natural} \,, 
\end{equation}
where 
\begin{equation}
q=\frac{f}{M_{\rm pl}}\,,\qquad 
\gamma=\frac{\Lambda^4}{M^3 M_{\rm pl}}\,.
\end{equation}
Note that in the limit $\phi \to 0$ one has ${\cal A} \to 0$.
We are mainly interested in the case where the initial displacement 
of the field $\phi_i$ satisfies the condition ${\cal A} (\phi_i)>1$.
This can be achieved for $4\gamma \gg q$ provided that 
$\phi_i$ is not very close to 0.

The field value $\phi_G$ at the transition (${\cal A}=1$) 
is given by 
\begin{equation}
\sin( \phi_G / f )=2q/\gamma \,.
\end{equation}
For the existence of $\phi_G$ we require that $2q<\gamma$, 
i.e., $M^3<\Lambda^4/(2f)$.
{}From Eq.~(\ref{phif2}) the field value $\phi_f$ at the end 
of inflation satisfies
\begin{equation}
\frac{1-\cos(\phi_f/f)}{1+\cos(\phi_f/f)}
=q^2 \left[ 1+\sqrt{1+\frac{4\gamma\sin(\phi_f/f)}{q}}
\right]\,.
\label{phif3}
\end{equation}
The transition from Galileon inflation to standard inflation 
occurs under the condition $\phi_G<\phi_f$. 
In the limit $\gamma \to 0$ one has $\cos(\phi_f/f)=(1-2q^2)/(1+2q^2)$, 
so that $\phi_f/f \to 0$ for $q \ll 1$.
This implies that, in the absence of the Galileon self-interaction, 
it is difficult to realize sufficient amount of inflation for $f \ll M_{\rm pl}$.
If the Galileon term is present with $\gamma \to \infty$, 
the field value $\phi_f$ can be close to $\pi f$ 
even for $f \ll M_{\rm pl}$.

\begin{figure}
\includegraphics[height=3.3in,width=3.2in]{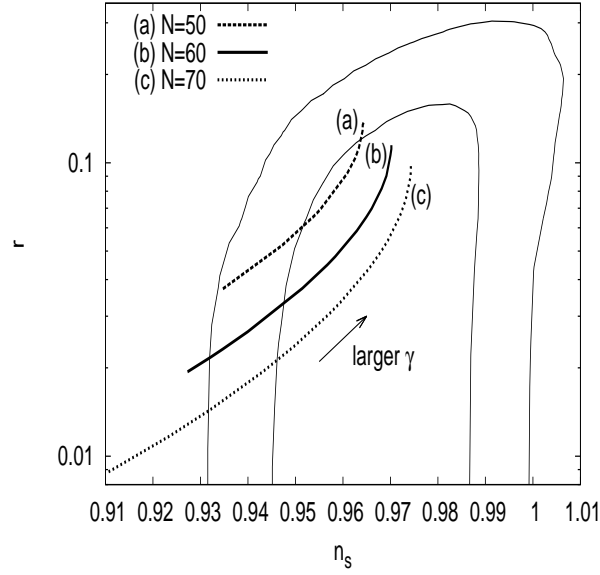}
\caption{\label{fig6}
Observational constraints on natural inflation with $f=0.1M_{\rm pl}$
in the presence of the term $G_3=X/M^3$
for three different values of $N$.
The parameter range of $\gamma=\Lambda^4/(M^3 M_{\rm pl})$ corresponds to  
$4.0\times10^5\leq \gamma \leq 1.0 \times 10^{11}$.
The $1\sigma$ and $2\sigma$ observational contours 
are the same as those in Fig.~\ref{fig3}.}
\end{figure}

\begin{figure}
\includegraphics[height=3.3in,width=3.2in]{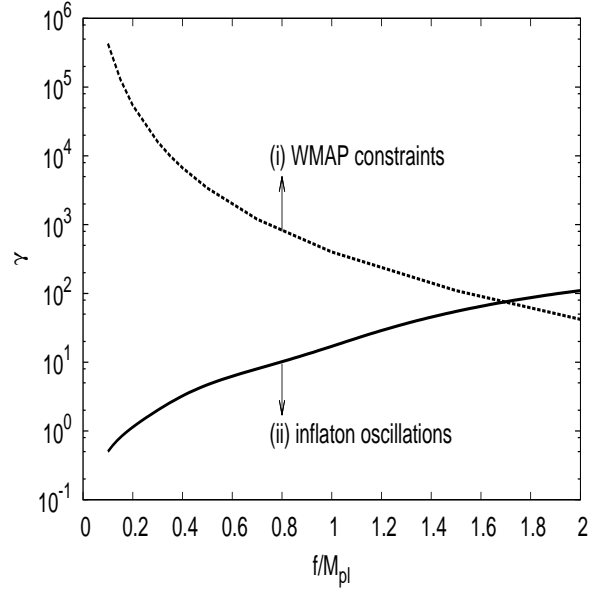}
\caption{\label{fig7}
The allowed values of $\gamma$ versus $f/M_{\rm pl}$ 
in natural inflation with the term $G_3=X/M^3$.
In the region (i) the model is within the $2\sigma$ observational 
contour in the ($n_s, r$) plane for $N=60$. 
In the region (ii) the coherent oscillation of inflaton occurs
during reheating. 
There is a viable parameter space only for $f/M_{\rm pl}>1.7$.}
\end{figure}

The slow-roll parameters $\epsilon_{\phi}$ and $\eta_{\phi}$
are given by 
\begin{equation}
\epsilon_{\phi}=\frac{1}{2q^2} \frac{\sin^2 (\phi/f)}
{[1+\cos(\phi/f)]^2}\,,\qquad
\eta_{\phi}=-\frac{1}{q^2} \frac{\cos (\phi/f)}
{1+\cos (\phi/f)}\,.
\end{equation}
In the limit $\gamma \to 0$, i.e., ${\cal A} \to 0$, 
the scalar spectral index is $n_s \simeq 1-6\epsilon_{\phi}+2\eta_{\phi}$.
When $q \ll 1$ inflation occurs in the region $\phi/f  \ll 1$, so that 
$|\eta_{\phi}| \simeq 1/(2q^2) \gg 1$.
This case is in contradiction with observations because
$n_s$ significantly deviates from $1$.
In another limit $\gamma \to \infty$ one has ${\cal A} \to \infty$
and hence the scalar spectral index (\ref{nsG3}) 
can be as close as 1 even for $q \ll 1$.
In this limit, inflation occurs in the regime close to the 
potential minimum ($\phi/f=\pi$).
Since the potential is approximately given by 
$V(\phi) \simeq (\Lambda^4/2f^2) (\phi-\pi f)^2$ in this regime, 
$n_s$ and $r$ are the same as those given in Eq.~(\ref{nsrga})
with $n=2$. Hence, for $\gamma \to \infty$, it follows that  
\begin{equation}
n_s=1-\frac{9}{5N+2}\,,\qquad
r=\frac{128 \sqrt{6}}{9(5N+2)}\,,
\label{nsrnatu}
\end{equation}
which give $n_s=0.970$ and $r=0.115$ for $N=60$.

In the intermediate regime characterized by $0<\gamma<\infty$ 
we need to evaluate $n_s$ and $r$ numerically
according to Eqs.~(\ref{nsG3}) and (\ref{rG3}).
For given values of $f$, $\Lambda$, and $M$, $\phi_f$
is known by solving Eq.~(\ref{phif3}).
Integrating Eq.~(\ref{efoldG3}) numerically, we can determine
the field $\phi$ in terms of the number of e-foldings $N$.
The WMAP normalization ${\cal P}_s=2.4 \times 10^{-9}$
at $N=60$ provides one constraint between the three 
parameters $f$, $\Lambda$, and $M$.
In other words, for a given $f$, the two parameters $\Lambda$
and $M$ are related to each other.

Let us first consider the case $f=0.1M_{\rm pl}$, i.e., $q=0.1$.
In Fig.~\ref{fig6}, we plot the theoretical values of $n_s$ and $r$ 
in the range $4.0\times10^5\leq \gamma\leq 1.0\times10^{11}$
for three different values of $N$, together with 
the $1 \sigma$ and $2 \sigma$ observational contours.
When $N = 60$ the model is within the $2\sigma$ contour 
provided that $\gamma >4.3\times10^{5}$.
In the limit $\gamma \to \infty$, $n_s$ and $r$ approach
the asymptotic values given in Eq.~(\ref{nsrnatu}).
This asymptotic case is within the 
$1\sigma$ contour for $N>55$.
For larger $\gamma$, however, the inflaton oscillation 
during reheating tends to be disturbed by the Galileon term.
After the inflaton velocity $\dot{\phi}$ changes its sign from positive 
to negative, the condition (\ref{concon}) can be violated for $\gamma \gg 1$.
Detailed numerical simulations show that 
the coherent oscillations of inflaton occur provided that $\gamma<0.5$.
Thus there is no viable parameter space of $\gamma$ 
satisfying both the WMAP bound and the successful reheating.
We also note that, if $\gamma$ is larger than 0.05, 
$c_s^2$ becomes negative.
Hence, for $\gamma>0.5$, the model is plagued by the reheating 
problem as well as the negative instability of scalar perturbations.

We also study the models of other values of $f$ ranging in the region 
$0.1<f/M_{\rm pl}<2$. In Fig.~\ref{fig7} we show the two kinds of constraints 
on the parameter $\gamma$ versus $f/M_{\rm pl}$.
Above the dotted line (i) the model is within the $2\sigma$ observational 
contour in the ($n_s, r$) plane, whereas under the solid line (ii)
the coherent oscillation of inflaton occurs during reheating.
For the compatibility of two constraints we require that $f$ 
is bounded to be 
\begin{equation}
f > 1.7\,M_{\rm pl}\,.
\end{equation}
Hence the problem of the super-Planckian values of $f$ in standard 
natural inflation is not circumvented by the Galileon term $G_3=X/M^3$.

\section{Theories with $G_3=0, G_4 \neq 0, G_5=0$}
\label{G4}

We proceed to the covariant Galileon theory (\ref{fi}) with 
\begin{equation}
c_3=0\,,\qquad c_4\neq 0\,,\qquad c_5=0\,.
\end{equation}
Since $c_4<0$ to avoid ghosts, we set $c_4=-1$ 
without loss of generality. {}From Eq.~(\ref{slowba2}) 
the quantity $\A=18 (H \dot{\phi}/M^3)^2$ satisfies 
the following relation 
\begin{equation}
\A (1+\A)^2=2V_{,\phi}^2/M^6\,.
\label{AG4}
\end{equation}
The field value $\phi_G$ at the transition from Galileon inflation
to standard inflation obeys
\begin{equation}
V_{,\phi}^2 (\phi_G)=2M^6\,.
\label{phiGG4}
\end{equation}

{}From Eq.~(\ref{Adef}) we have $\delta_4=\A \delta_X/18$.
Using Eq.~(\ref{delXslow}), the scalar power spectrum (\ref{Psf})
can be written as 
\begin{equation}
{\cal P}_s=\frac{V^3}{12\pi^2 M_{\rm pl}^6 V_{,\phi}^2}
\frac{(1+\A)^2 (1+3\A)^{1/2}}{(1+13\A/9)^{3/2}}\,.
\end{equation}
Taking the time derivative of Eq.~(\ref{AG4}) and 
making use of Eq.~(\ref{slowba2}), it follows that 
$\dot{\A}/H=-2\eta_{\phi} \A/(1+3\A)$.
Then the scalar spectral index is given by 
\begin{equation}
n_s-1=-\frac{6\epsilon_{\phi}}{1+\A}
+\frac{2\eta_{\phi}}{1+3\A/2}
\left[ 1-\frac{3\A (5+8\A)}{2(1+3\A)^2(9+13\A)} 
\right]\,.
\label{nsG4}
\end{equation}
The tensor-to-scalar ratio (\ref{rgene}) reads
\begin{equation}
r=16\epsilon_{\phi} \frac{(1+13\A/9)^{3/2}}
{(1+\A)^2 (1+3\A)^{1/2}}\,.
\label{rG4}
\end{equation}
In the limit $\A \to \infty$ we have 
$n_s-1 \simeq -6\epsilon_{\phi}/\A+4\eta_{\phi}/(3\A)$ and 
$r \simeq 208 \sqrt{39}\, \epsilon_{\phi}/(81\A)$.

In the following we focus on the potential $V(\phi)=\lambda \phi^n/n$
of chaotic inflation.
{}From Eq.~(\ref{phiGG4}) the field value at the transition is
\begin{equation}
\phi_G=\left( 2M^6/\lambda^2 \right)^{1/[2(n-1)]}\,.
\end{equation}
The condition under which the transition occurs during inflation 
corresponds to $\epsilon(\phi_G)<1$, which translates into
\begin{equation}
M>2^{(1-2n)/6} n^{(n-1)/3} M_{\rm pl}^{(n-1)/3}
\lambda^{1/3}\,.
\label{G4Mcon}
\end{equation}
%

\begin{figure}
\includegraphics[height=3.1in,width=3.2in]{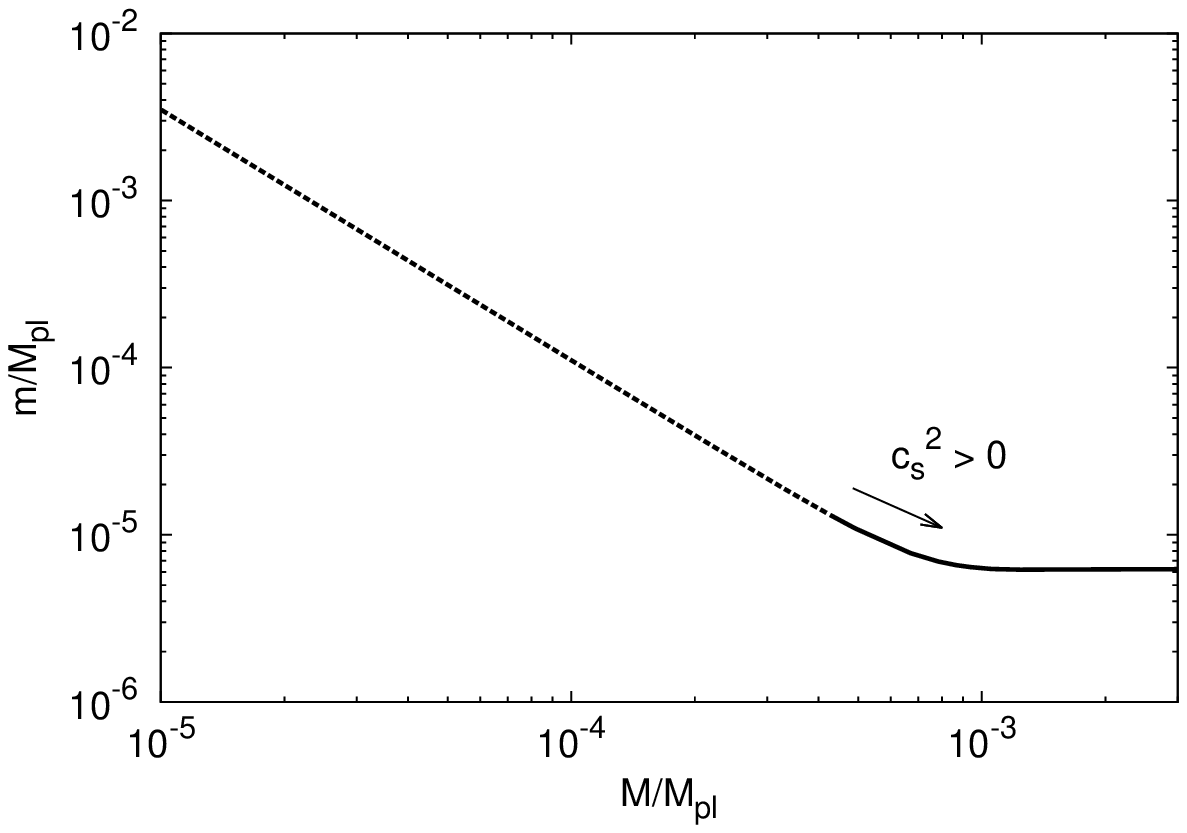}
\includegraphics[height=3.1in,width=3.2in]{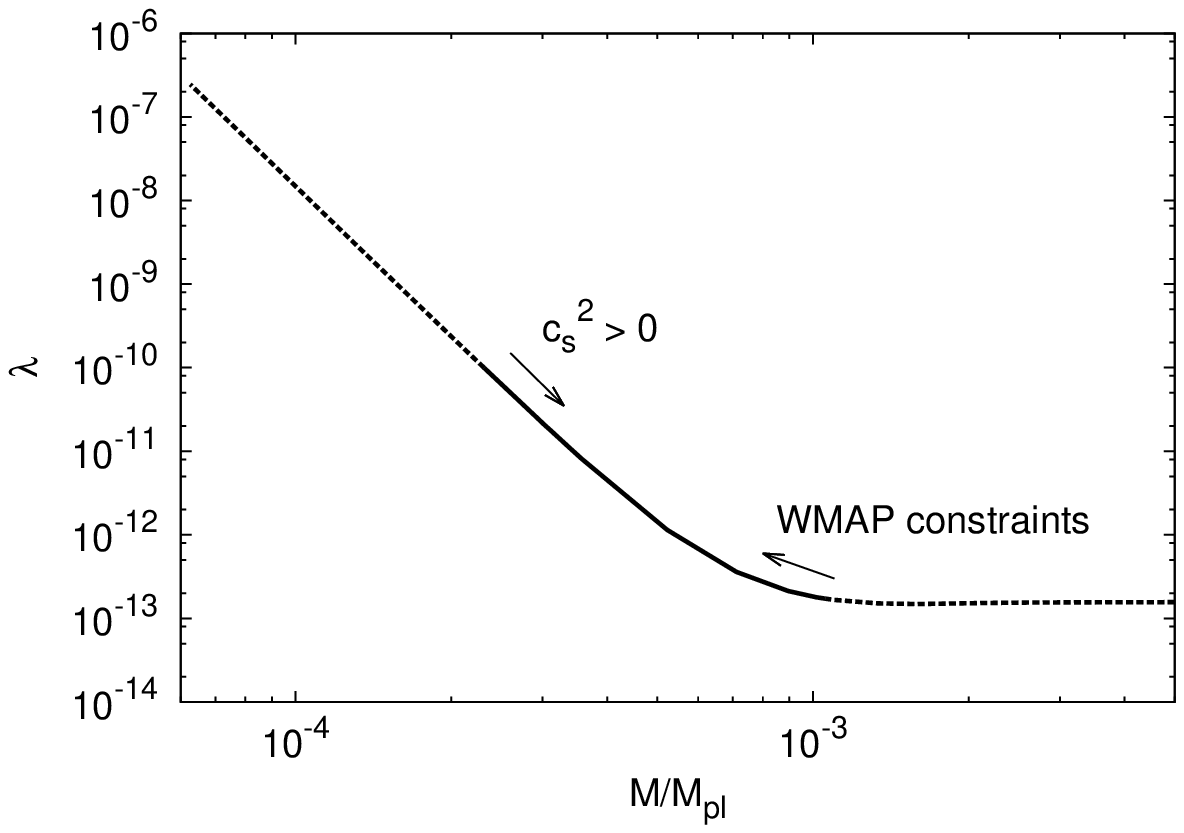}
\caption{\label{fig8}
The parameter space satisfying the normalization 
${\cal P}_s=2.4 \times 10^{-9}$ at $N=60$ for the quadratic potential 
$V(\phi)=m^2 \phi^2/2$ (left) and for the quartic potential 
$V(\phi)=\lambda \phi^4/4$ (right) in the 
presence of the term $G_4=X^2/M^6$.
The solid lines correspond to the regions in which $c_s^2$
is positive and the observational constraints of $n_s$ and $r$ are satisfied.}
\end{figure}

\begin{figure}
\includegraphics[height=3.1in,width=3.3in]{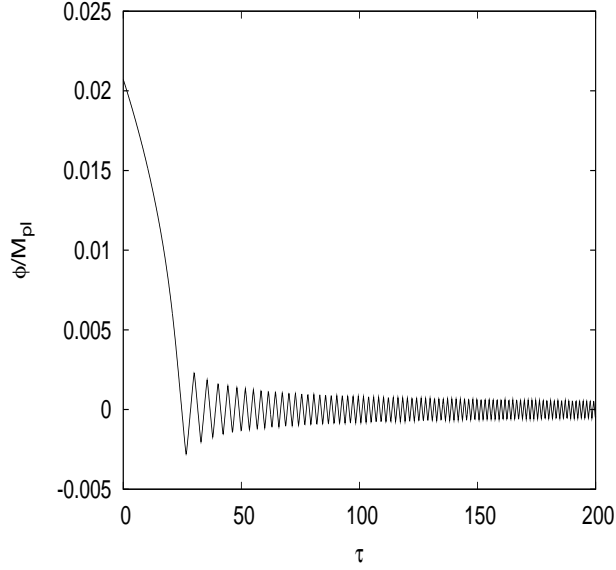}
\caption{\label{fig9}
Evolution of the field $\phi$ for 
the quartic potential $V(\phi)=\lambda \phi^4/4$ with 
$\lambda=0.1$ in the presence of the term 
$G_4=X^2/M^6$ with
$M=7.3 \times 10^{-6} M_{\rm pl}$.
The initial conditions are chosen to be 
$x_i=2.07 \times 10^{-2}$, 
$y_i=4.69 \times 10^{-4}$, and 
$z_i=5.49$ at $N=60$.
}
\end{figure}

\begin{figure}
\includegraphics[height=3.1in,width=3.3in]{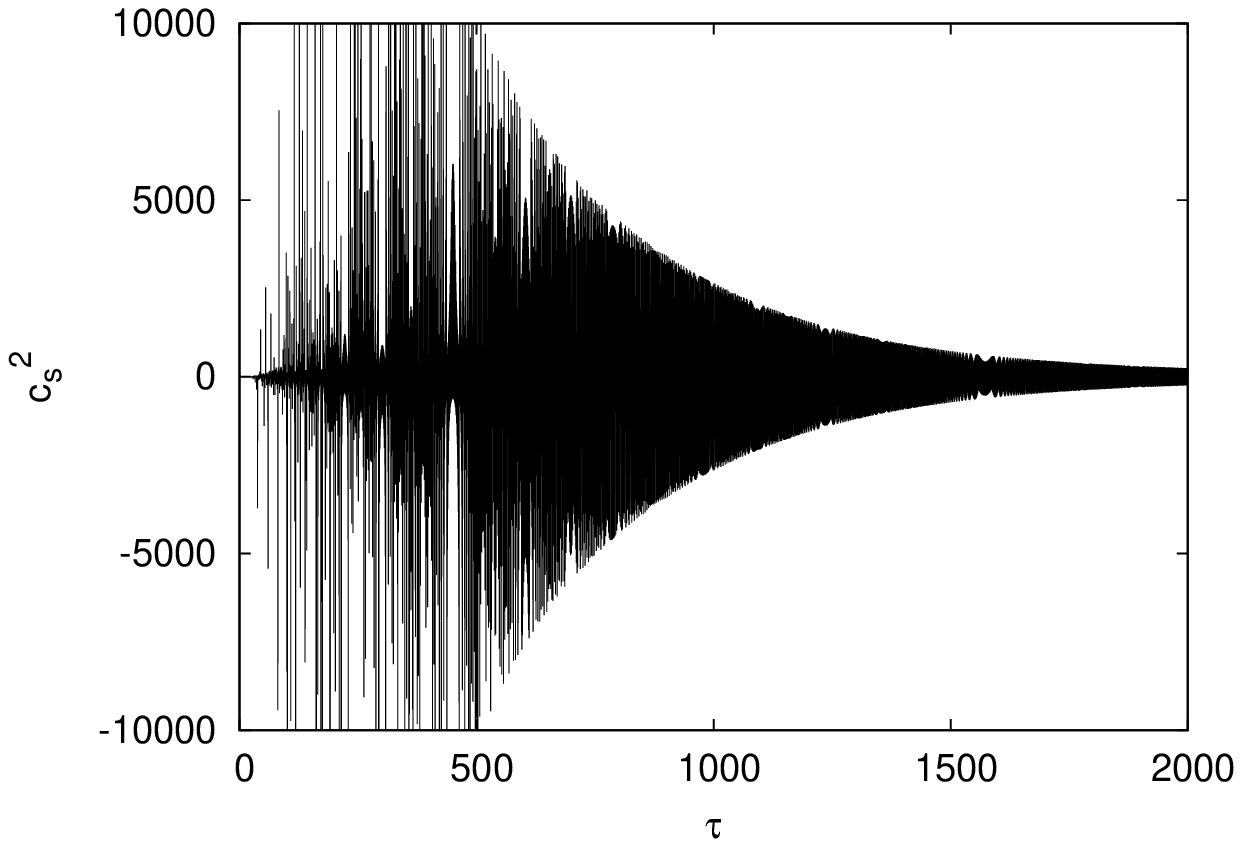}
\includegraphics[height=3.1in,width=3.3in]{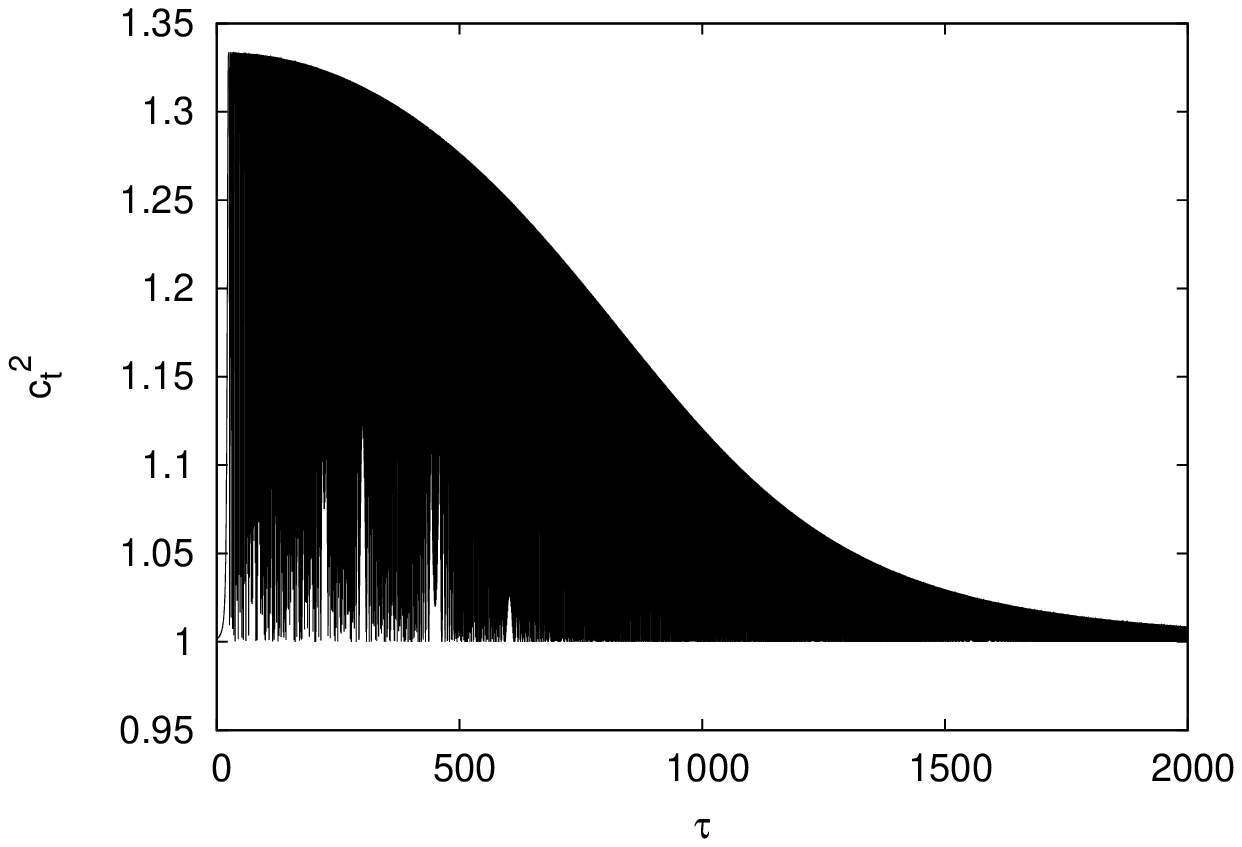}
\caption{\label{fig10}
The same as Fig.~\ref{fig9}, but for the evolution 
of $c_s^2$ (left) and $c_t^2$ (right).
}
\end{figure}

\begin{figure}
\includegraphics[height=3.1in,width=3.1in]{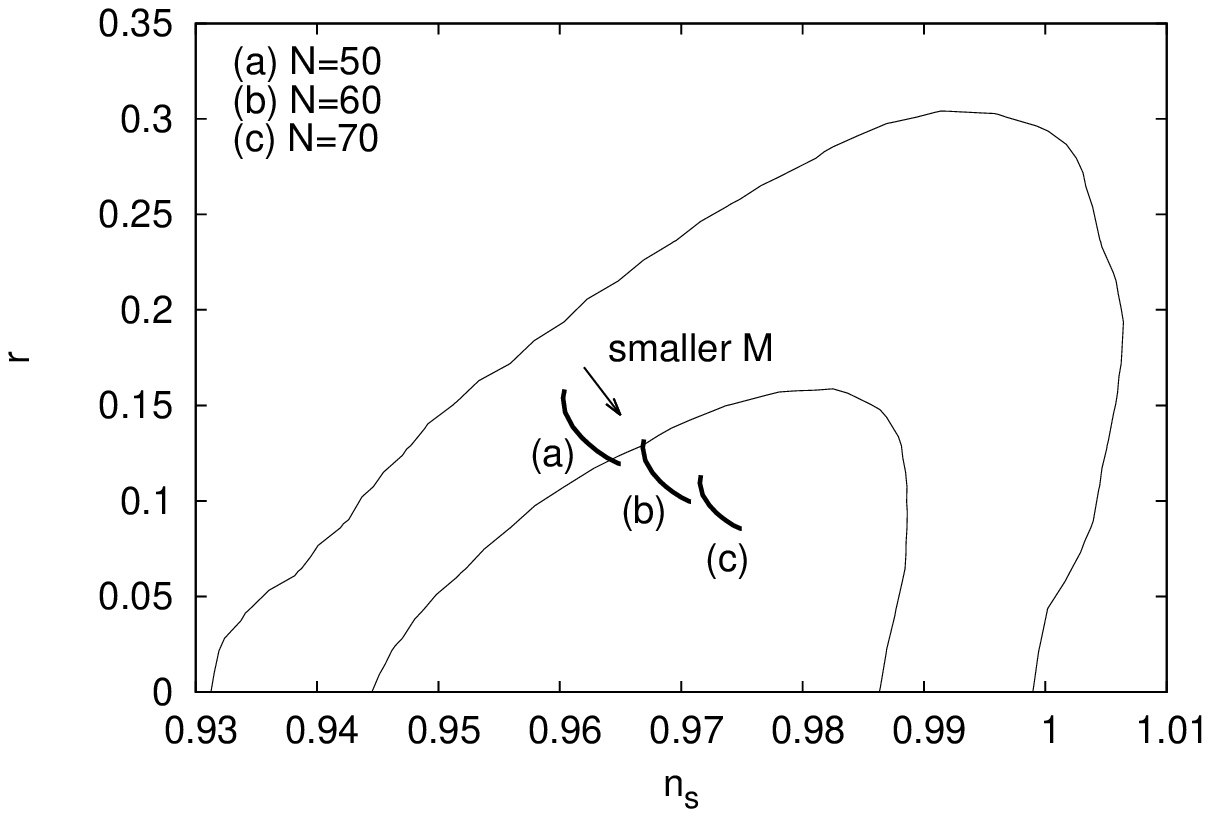}
\includegraphics[height=3.1in,width=3.1in]{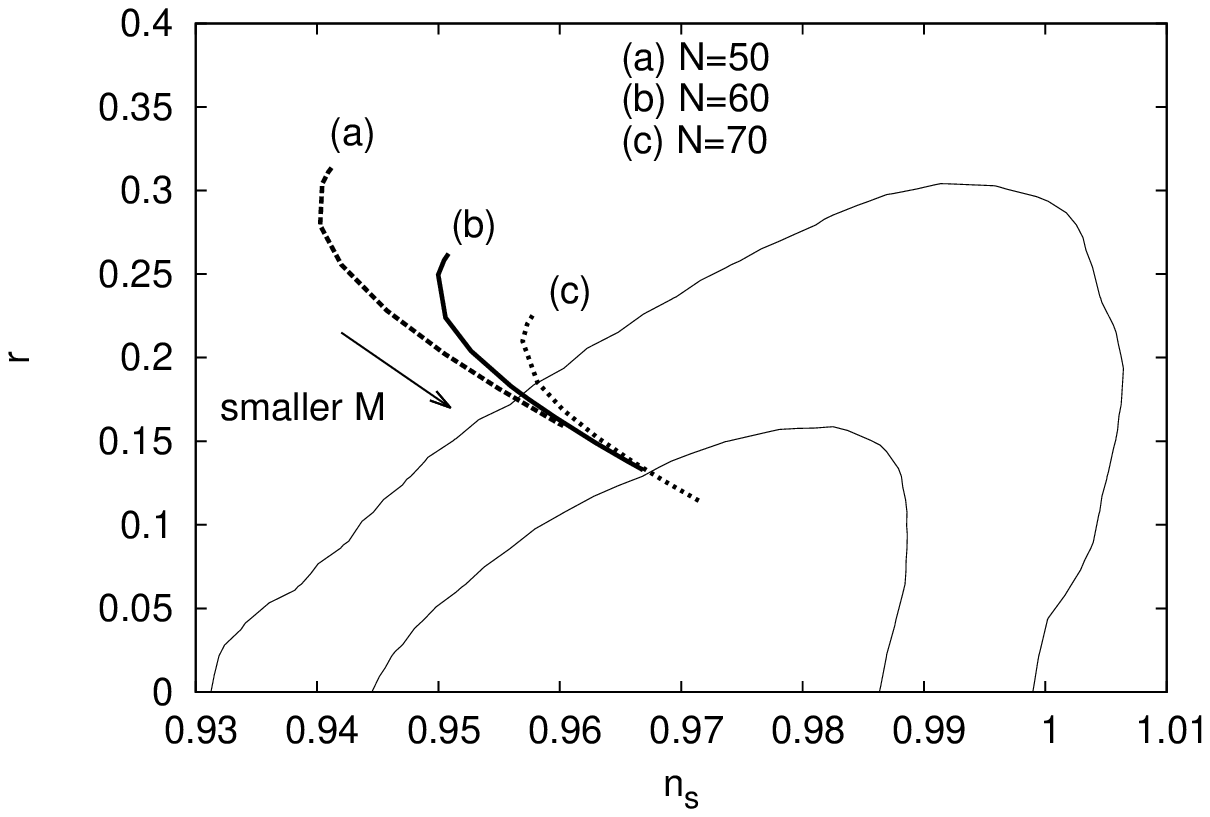}
\caption{\label{fig11}
Observational constraints in the $(n_s, r)$ plane with the 
numbers of e-foldings $N=50\,, 60\,, 70$ for the quadratic potential 
$V(\phi)=m^2 \phi^2/2$ (left) and for the quartic potential 
$V(\phi)=\lambda \phi^4/4$ (right) in the 
presence of the term $G_4=X^2/M^6$.
The $1\sigma$ and $2\sigma$ observational contours 
are the same as those shown in Fig.~\ref{fig3}.}
\end{figure}

Let us consider the case in which the condition $\A \gg 1$
is satisfied during the whole stage of inflation.
Then we have $54 (H \dot{\phi})^3 \simeq -M^6 V_{,\phi}$
from Eq.~(\ref{slowba2}).
The end of inflation is characterized by the condition 
$\epsilon(\phi_f) \simeq \epsilon_{\phi} (\phi_f)/\A=1$,
which gives 
$\phi_f=[ n^3 M_{\rm pl}^3 M^3/(4\lambda)]^{1/(n+2)}$.
The number of e-foldings (\ref{Ndef}) is related to 
the field value $\phi$ during inflation, as 
\begin{equation}
\phi^{2(n+2)/3} \simeq \frac{nM_{\rm pl}^2 M^2}{6(2\lambda^2)^{1/3}}
\left[ 4(n+2)N+3n \right]\,.
\label{phiG4}
\end{equation}
The WMAP normalization ${\cal P}_s=2.4 \times 10^{-9}$ 
at $N=60$ provides the relation
\begin{equation}
\lambda^2 \frac{M^{3n}}{M_{\rm pl}^{n+8}}
=\frac{2^{3n+2}\,n^{n/2+4}}{39^{n/2+1}}
\frac{(1.8 \times 10^{-6})^{n+2}}{(81n+160)^{(5n+4)/2}}\,.
\end{equation}
Substituting Eq.~(\ref{phiG4}) into Eqs.~(\ref{nsG4}) and (\ref{rG4})
in the regime $\A \gg 1$, it follows that 
\begin{equation}
n_s=1-\frac{2(5n+4)}{4(n+2)N+3n}\,,\qquad
r=\frac{208 \sqrt{39}}{27} 
\frac{n}{4(n+2)N+3n}\,.
\label{nsrG4lim}
\end{equation}
For $n=4$ and $N=60$, for example,  
$n_s=0.967$ and $r=0.133$.
The tensor-to-scalar ratio is smaller than that 
studied in Sec.~\ref{G3} in the regime $\A \gg 1$.

In another limit ${\cal A} \ll 1$, we have the same relations as those 
given in Eqs.~(\ref{WMAP2}) and (\ref{nsrga2}). 
In the intermediate regime between ${\cal A} \gg 1$ and ${\cal A} \ll 1$
we need to solve the background equations (\ref{contsrainteq})-(\ref{eom2}) 
numerically in order to find the values of $n_s$ and $r$ as well as
the relation between $\lambda$ and $M$ from 
the WMAP normalization.
In Fig.~\ref{fig8} we show the parameter space for the two 
potentials $V(\phi)=m^2 \phi^2/2$ and $V(\phi)=\lambda \phi^4/4$
satisfying the WMAP normalization at $N=60$. 
In the two asymptotic regimes ${\cal A} \gg 1$ and ${\cal A} \ll 1$, 
the analytic estimation given above agrees 
well with the numerical results.

Unlike the case of the coupling $G_3=c_3 X/M^3$, the term 
$1+54 H^2 \dot{\phi}^2/M^6$ in front of $\ddot{\phi}$ in Eq.~(\ref{E3})
remains positive even if $\dot{\phi}$ changes its sign.
Numerically we confirmed that the determinant $\Delta$ defined in 
Eq.~(\ref{deter}) does not cross 0 even for the mass $M$
much smaller than the r.h.s. of Eq.~(\ref{G4Mcon}).
In Fig.~\ref{fig9} we show the field evolution during reheating 
for the quartic potential $V(\phi)=\lambda \phi^4/4$
with $\lambda=0.1$ and $M=7.3 \times 10^{-6} M_{\rm pl}$.
In fact the coherent oscillation of inflaton is 
not disturbed by the dominance 
of the term $G_4=X^2/M^6$.
In this case, however, the scalar propagation speed squared
oscillates significantly between largely negative and positive values
(see the left panel of Fig.~\ref{fig10}).
This leads to the strong enhancement of scalar perturbations for 
the modes inside the Hubble radius during reheating.
While this instability does not directly affect the evolution of 
large-scale density perturbations relevant to CMB, the rapid growth of 
perturbations can invalidate the analysis without 
the backreaction of created particles after some stage of reheating \cite{KLS}.
Our numerical simulations without the backreaction effect show that both 
$c_s^2$ and $c_t^2$ finally approach 1 with oscillations. 
The tensor propagation speed is superluminal 
during most stages of inflation and reheating, but it
does not enter the region $c_t^2<0$ (see the right 
panel of Fig.~\ref{fig10}).

It remains to see how the created particles can 
change the evolution of $\phi$, $c_s^2$, and $c_t^2$ at the late stage 
of reheating. This is beyond the scope of our paper, since
nonlinear lattice simulations (along the line of Refs.~\cite{lattice}) 
are required to deal with such a problem properly.

For larger values of $M$, the instability associated with 
negative $c_s^2$ tends to be less significant.
For the power-law potential $V(\phi)=\lambda \phi^n/n$
we find that $c_s^2$ remains positive for
\begin{eqnarray}
& & M>4.3\times10^{-4}M_{\rm pl} \,\qquad(\text{for $n=2$})\,,
\label{con1G40} \\
& & M>2.3\times10^{-4}M_{\rm pl} \,\qquad  (\text{for $n=4$})\,,
\label{con1G4}
\end{eqnarray}
respectively. 
The regions in which these conditions are satisfied 
are shown as solid curves in Fig.~\ref{fig8}.
In Fig.~\ref{fig11} we plot the theoretical values of $n_s$ 
and $r$ for $n=2$ and $n=4$ as a function of $M$.
Even for the lower bounds of Eqs.~(\ref{con1G40}) 
and (\ref{con1G4}), $n_s$ and $r$ are close 
to the values (\ref{nsrG4lim}) corresponding to the 
limit ${\cal A} \gg 1$.
For the quadratic potential the presence of the 
term $G_4=X^2/M^6$ leads to better compatibility 
with the WMAP data 
(see the left panel of Fig.~\ref{fig11}).
In the case of the quartic potential the model is 
within the $2\sigma$ observational contour 
under the condition
\begin{equation}
M<1.1\times10^{-3}M_{\rm pl}\,,
\label{con2G4}
\end{equation}
for $N=60$.
{}From Fig.~\ref{fig8} this condition translates into
\begin{equation}
\lambda >1.7\times10^{-13}\,.
\label{lam_G4}
\end{equation}
If we demand the condition (\ref{con1G4}) for the 
avoidance of negative values of $c_s^2$,
the self coupling is bounded to be
$\lambda<9.9 \times 10^{-11}$.
Recall that we do not have a constraint coming from 
the absence of inflaton oscillations.

\section{Theories with $G_3=0, G_4=0, G_5 \neq 0$}
\label{G5}

Finally we study the covariant Galileon theory (\ref{fi}) with
\begin{equation}
c_3=0\,,\qquad c_4=0\,,\qquad c_5 \neq 0\,.
\end{equation}
{}From Eq.~(\ref{slowba2}) the quantity 
$\A=15c_5 H^3 \dot{\phi}^3/M^9$ satisfies 
\begin{equation}
\A (1+\A)^3=-5c_5 V_{,\phi}^3/(9M^9)\,.
\label{AG5}
\end{equation}
The field value $\phi_G$ at the transition from
Galileon inflation to standard inflation is determined by 
\begin{equation}
V_{,\phi}^3 (\phi_G)=-72M^9/(5c_5)\,.
\label{phiGG5}
\end{equation}
Since $\delta_5=\A \delta_X/5$, 
the scalar power spectrum (\ref{Psf}) reduces to 
\begin{equation}
{\cal P}_s=\frac{V^3}{12\pi^2 M_{\rm pl}^6 V_{,\phi}^2}
\frac{(1+\A)^2 (1+4\A)^{1/2}}{(1+8\A/5)^{3/2}}\,.
\end{equation}
On using the relation $\dot{\A}/H=-3\eta_{\phi} \A/(1+4\A)$,
the scalar spectral index is expressed as 
\begin{equation}
n_s-1=-\frac{6\epsilon_{\phi}}{1+\A}
+\frac{2\eta_{\phi}}{1+8\A/5}
\left[ 1-\frac{9\A}{5(1+4\A)^2} 
\right]\,.
\label{nsG5}
\end{equation}
The tensor-to-scalar ratio (\ref{rgene}) reads
\begin{equation}
r=16\epsilon_{\phi} \frac{(1+8\A/5)^{3/2}}
{(1+\A)^2 (1+4\A)^{1/2}}\,.
\label{rG5}
\end{equation}

Let us focus on the power-law potential $V(\phi)=\lambda \phi^n/n$.
We assume that inflation occurs in the regime 
$\phi>0$ with the coefficient $c_5=-1$ (under which 
the condition $c_5 \dot{\phi}>0$ is satisfied).
Then the field value at the transition is
\begin{equation}
\phi_G=[72 M^9 /(5\lambda^3)]^{1/[3(n-1)]}\,.
\end{equation}
The condition under which the transition occurs during inflation is 
\begin{equation}
M>0.94 \cdot 2^{-n/3} n^{(n-1)/3} M_{\rm pl}^{(n-1)/3}
\lambda^{1/3}\,.
\label{G5Mcon}
\end{equation}

If $\A \gg 1$ during the whole stage of inflation, 
the field value at the end of inflation can be estimated as
$\phi_f=[ (9/5)^{1/4} n^2 M_{\rm pl}^2 M^{9/4}/(2\lambda^{3/4})]^{4/(3n+5)}$.
The field $\phi$ is related to the number of e-foldings $N$, as
\begin{equation}
\phi^{(3n+5)/4} \simeq \frac{nM_{\rm pl}^2}{4\lambda^{3/4}}
\left( \frac{9M^9}{5} \right)^{1/4}
\left[ (3n+5)N+2n \right]\,.
\label{phiG5}
\end{equation}
{}From the WMAP normalization ${\cal P}_s=2.4 \times 10^{-9}$ 
at $N=60$ it follows that  
\begin{equation}
\lambda^5 \frac{M^{9n}}{M_{\rm pl}^{4(n+5)}}
=1.33 \times 10^{-7} \cdot (4.44 \times 10^{-7})^{n/2} \cdot
(1.36 \times 10^{-5})^{3n+5}
\frac{n^{2(n+5)}}{(91n+150)^{7n+5}}\,.
\end{equation}
{}From Eqs.~(\ref{nsG5}) and (\ref{rG5}) the asymptotic values 
of $n_s$ and $r$ in the regime ${\cal A} \gg 1$ are
\begin{equation}
n_s=1-\frac{7n+5}{(3n+5)N+2n}\,,\qquad
r=\frac{256 \sqrt{10}}{25} 
\frac{n}{(3n+5)N+2n}\,.
\end{equation}
If $n=4$ and $N=60$, for example, 
$n_s=0.968$ and $r=0.126$.
The tensor-to-scalar ratio is slightly smaller than that for
the coupling $G_4=X^2/M^6$.
In the regime ${\cal A} \ll 1$ the relations (\ref{WMAP2})
and (\ref{nsrga2}) also hold for the coupling $G_5=-3X^2/M^9$.
In the intermediate regime between ${\cal A} \gg 1$ and ${\cal A} \ll 1$
we resort to the numerical analysis to derive the relation between 
$M$ and $\lambda$ from the WMAP normalization as well as to 
evaluate the observables $n_s$ and $r$.

In order to avoid that the term in front of $\ddot{\phi}$ in Eq.~(\ref{E3}) 
becomes negative, we require that 
\begin{equation}
1+120H^3 X c_5 \dot{\phi}/M^9>0\,.
\end{equation}
For smaller $M$ this condition can be violated during reheating because 
of the sign change of $\dot{\phi}$ (as it happens for the coupling 
$G_3=c_3 X/M^3$). 
Note that this also leads to the divergence 
of Eqs.~(\ref{eom2}) and (\ref{eom3}) through the crossing at $\Delta=0$.
For the quadratic and quartic potentials we find that the inflaton 
oscillations occur under the conditions
\begin{eqnarray}
& &M>2.7\times10^{-4}M_{\rm pl} \qquad (\text{for $n=2$})\,,
\label{con1G5d} \\
& &M>1.5\times10^{-4}M_{\rm pl} \qquad (\text{for $n=4$})\,,
\label{con1G5}
\end{eqnarray}
respectively. The instability associated with negative values of
$c_s^2$ tends to be stronger for smaller $M$.
The conditions under which $c_s^2$ remains positive 
are given by 
\begin{eqnarray}
& &M>4.0\times10^{-4}M_{\rm pl} \qquad (\text{for $n=2$})\,, 
\label{concG5d}\\
& &M>2.9\times10^{-4}M_{\rm pl} \qquad (\text{for $n=4$})\,,
\label{concG5}
\end{eqnarray}
respectively.
The similar lower bounds on $M$ ($\gtrsim 10^{-4}M_{\rm pl}$) 
also follow from Eq.~(\ref{G5Mcon}). 

In the presence of the coupling $G_5=-3X^2/M^9$ 
the tensor-to-scalar ratio gets smaller relative to that 
in standard inflation, 
so that the quadratic potential $V(\phi)=m^2 \phi^2/2$ is 
compatible with the current observational data.  
For the quartic potential $V(\phi)=\lambda \phi^4/4$
the model is within the $2\sigma$ observational contour 
in the $(n_s,r)$ plane under the condition
\begin{equation}
M<8.6\times 10^{-4} M_{\rm pl}\,,
\label{con2G5}
\end{equation}
for $N=60$.
Translating the conditions (\ref{con1G5}) and (\ref{con2G5})
in terms of the parameter $\lambda$, it follows that 
\begin{equation}
1.6\times10^{-13}<\lambda <2.6 \times10^{-9}\,.
\label{lam_G5}
\end{equation}
As in the case of the coupling $G_3=-X/M^3$, the self coupling $\lambda$ 
is required to be very much smaller than unity.

\section{Conclusions}
\label{Conclusions}

We have studied the viability of potential-driven Galileon inflation 
described by the action (\ref{kinfaction}).
We mainly focused on the covariant Galileon theory in which the functions 
$G_i$ ($i=3,4,5$) are given by Eq.~(\ref{Gi}) with the choice (\ref{fi}).
The Galileon self-interactions generally lead to the slow down for the 
evolution of the field, which allows the possibility to accommodate
steep inflaton potentials.
In Ref.~\cite{Kamada:2010qe}, for example, it was suggested that even
the Higgs potential $V(\phi)=\lambda \phi^4/4$ with $\lambda \sim 0.1$ 
can be consistent with the observed CMB temperature anisotropies
because of the presence of the term $G_3=c_3 X/M^3$.

The dominance of the Galileon self-interactions relative to the standard
kinetic term $X$ can modify the dynamics of reheating after inflation.
In order to clarify this issue, we numerically solved the background equations
(\ref{contsrainteq})-(\ref{eom3}) for several different inflaton potentials.
We found that, depending on the couplings $G_i$ ($i=3,4,5$) and 
their associated mass scales $M$, there is no oscillatory regime of inflaton.
Moreover the dominance of the Galileon terms generally gives rise
to the negative scalar propagation speed squared $c_s^2$ during 
reheating, which leads to the instability of small-scale 
density perturbations.

For the theories where the covariant Galileon 
term $G_3=c_3 X/M^3$ is present, 
we found that the system does not enter the oscillatory regime
of inflaton after the field velocity $\dot{\phi}$ changes its sign around 
the onset of reheating.
This corresponds to the violation of the condition (\ref{concon}), 
which is related to the crossing of the determinant $\Delta$ in 
Eq.~(\ref{deter}) at 0.
The latter leads to the divergence of 
the background equations (\ref{eom2}) and (\ref{eom3}).
For the  potentials $V(\phi)=\lambda \phi^n/n$ 
the coherent oscillation of inflaton occurs for
$M>2.5 \times 10^{-4}M_{\rm pl}$ ($n=2$) and 
$M>9.5\times10^{-5}M_{\rm pl}$ ($n=4$).
When $n=4$ this constraint translates into $\lambda <3.1\times10^{-10}$, 
which is much smaller than the coupling constant $\lambda\sim0.1$ 
of the Higgs boson. 
In the presence of the term $G_3=c_3 X/M^3$ the quartic
potential $V(\phi)=\lambda \phi^4/4$ is within the 2$\sigma$
observational contour in the $(n_s,r)$ plane 
for $M<7.7 \times 10^{-4} M_{\rm pl}$.
Taking into account this constraint, the self coupling is 
bounded to be $3.4 \times 10^{-13}<\lambda<3.1 \times 10^{-10}$.
We also found that $c_s^2$ remains positive
under the condition $M>1.7 \times 10^{-4}M_{\rm pl}$,
which provides even the stronger upper bound $\lambda<3.0 \times 10^{-11}$.
We extended our analysis to the generalized Galileon term $G_3=c_3 \phi X/M^4$
with the potential $V(\phi)=\lambda \phi^4/4$
and derived the bound $\lambda<2.7 \times 10^{-8}$ for 
successful reheating.

In the presence of the term $G_3=c_3 X/M^3$ we studied the case 
of natural inflation described by the potential $V(\phi)=\Lambda^4[1+\cos(\phi/f)]$ 
as well. While this potential can be compatible with the observed CMB anisotropies
for $\gamma=\Lambda^4/(M^3 M_{\rm pl}) \gg 1$ even in the regime 
$f \ll M_{\rm pl}$, there is no oscillatory regime under the condition 
$\gamma \gg 1$. For the compatibility of two constraints, we found that 
$f$ needs to be larger than $1.7 M_{\rm pl}$.
Hence the super-Planckian problem of the symmetry breaking scale in 
standard inflation ($f>3.5M_{\rm pl}$) is not improved significantly.

For the Galileon coupling $G_4=-c_4 X^2/M^6$ the scalar ghost 
is absent for $c_4<0$, in which case the sign 
change of the determinant $\Delta$ in Eq.~(\ref{deter}) can be avoided.
In fact, we numerically confirmed that the oscillation of inflaton occurs
even for small $M$ corresponding to the large self coupling 
$\lambda \sim 0.1$ of the quartic potential $V(\phi)=\lambda \phi^4/4$.
On the other hand, for such small values of $M$, the scalar propagation 
speed squared $c_s^2$ heavily oscillates between largely negative and 
positive values (see the left panel of Fig.~\ref{fig10}).
This leads to the rapid growth of scalar perturbations for the modes
inside the Hubble radius during reheating, which can invalidate 
the analysis without taking into account 
the backreaction of created particles.
For the potentials $V(\phi)=\lambda \phi^n/n$ 
the conditions for the avoidance of this negative instability
are given by 
$M>4.3 \times 10^{-4} M_{\rm pl}$ ($n=2$) and 
$M>2.3 \times 10^{-4} M_{\rm pl}$ ($n=4$).
Taking into account this condition, the quartic potential 
$V(\phi)=\lambda \phi^4/4$ is compatible with the current CMB 
observations for $1.7 \times 10^{-13}<\lambda<9.9 \times 10^{-11}$.

In the case of the Galileon coupling $G_5=3c_5 X^2/M^9$ the sign change 
of $\Delta$ can occur for small $M$, as it happens for the coupling 
$G_3=c_3 X/M^3$. For the potentials $V(\phi)=\lambda \phi^n/n$ 
the inflaton oscillations occur for 
$M>2.7 \times 10^{-4} M_{\rm pl}$ ($n=2$) and 
$M>1.5 \times 10^{-4} M_{\rm pl}$ ($n=4$).
Using the latter bound, the quartic potential is consistent with 
the CMB observations for $1.6 \times 10^{-13}<\lambda<2.6 \times 10^{-9}$.
We also found that the instability associated with negative $c_s^2$
is present for small $M$, which puts even severer upper bounds on $\lambda$.

Compared to the models of non-minimal field derivative couplings to the 
Einstein tensor \cite{Germani:2010gm,Tsujikawa:2012mk},
the allowed parameter space of potential-driven Galileon inflation is 
more severely constrained because of the modified dynamics of reheating.
We note, however, that there are some viable parameter spaces even for the 
quartic potential $V(\phi)=\lambda \phi^4/4$ due to the presence of
the Galileon terms. It will be of interest to see whether future observations 
such as PLANCK \cite{Planck:2006aa} can place tighter constraints on 
such inflationary scenarios.

\section*{ACKNOWLEDGEMENTS}
J.\, O. and S.\,T. are supported by the 
Scientific Research Fund of the JSPS
(Nos.~23\,$\cdot$\,6781  and 24540286).
S.\,T. also thanks financial support from 
Scientific Research on Innovative Areas (No.~21111006). 


\begin{thebibliography}{10}

\bibitem{infpapers} 
A.~A.~Starobinsky,
Phys.\ Lett.\ B {\bf 91}, 99 (1980);
D.~Kazanas,
Astrophys.\ J.\  {\bf 241} L59 (1980);
K.~Sato, Mon.\ Not.\ R.\ Astron.\ Soc. {\bf 195}, 
467 (1981);
A.~H.~Guth,
Phys.\ Rev.\ D {\bf 23}, 347 (1981).

\bibitem{infper}
V.~F.~Mukhanov and G.~V.~Chibisov,
JETP Lett.\  {\bf 33}, 532 (1981);
A.~H.~Guth and S.~Y.~Pi,
Phys.\ Rev.\ Lett.\  {\bf 49}, 1110 (1982);
S.~W.~Hawking,
Phys.\ Lett.\ B {\bf 115}, 295 (1982);
A.~A.~Starobinsky,
Phys.\ Lett.\ B {\bf 117}, 175 (1982).

\bibitem{COBE}
G.~F.~Smoot {\it et al.},
Astrophys.\ J.\  {\bf 396}, L1-L5 (1992).

\bibitem{WMAP1}
D.~N.~Spergel {\it et al.}  [WMAP Collaboration],
Astrophys.\ J.\ Suppl.\  {\bf 148}, 175 (2003).

\bibitem{review}
J.~E.~Lidsey {\it et al.},
Rev.\ Mod.\ Phys.\  {\bf 69}, 373 (1997);
D.~H.~Lyth and A.~Riotto,
Phys.\ Rept.\  {\bf 314}, 1 (1999);
A.~D.~Linde,
 {\it ``Particle physics and inflationary cosmology,''}
Chur, Switzerland: Harwood (1990) 362 page 
(Contemporary concepts in physics, 5) [hep-th/0503203];
B.~A.~Bassett, S.~Tsujikawa and D.~Wands,
Rev.\ Mod.\ Phys.\  {\bf 78}, 537 (2006).
  
\bibitem{Linde:1983gd}
A.~D.~Linde,
Phys.\ Lett.\ B {\bf 129}, 177 (1983).
  
\bibitem{Komatsu:2010fb}
E.~Komatsu {\it et al.}  [WMAP Collaboration],
Astrophys.\ J.\ Suppl.\  {\bf 192}, 18 (2011).  

\bibitem{Amsler:2008zzb} 
C.~Amsler {\it et al.}  [Particle Data Group Collaboration],
Phys.\ Lett.\ B {\bf 667}, 1 (2008).

\bibitem{Nakayama} 
K.~Nakayama and F.~Takahashi,
JCAP {\bf 1011}, 009 (2010).
  
\bibitem{DeTavakol} 
A.~De Felice, S.~Tsujikawa, J.~Elliston and R.~Tavakol,
JCAP {\bf 1108}, 021 (2011).
  
 \bibitem{Varun} 
S.~Unnikrishnan, V.~Sahni and A.~Toporensky,
arXiv:1205.0786 [astro-ph.CO]. 
  
\bibitem{Maeda}
T.~Futamase and K.~-i.~Maeda,
Phys.\ Rev.\ D {\bf 39}, 399 (1989);
R.~Fakir and W.~G.~Unruh,
Phys.\ Rev.\ D {\bf 41}, 1783 (1990).

\bibitem{Bezrukov:2007ep} 
F.~L.~Bezrukov and M.~Shaposhnikov,
Phys.\ Lett.\ B {\bf 659}, 703 (2008).

\bibitem{Salopek}
D.~S.~Salopek, J.~R.~Bond and J.~M.~Bardeen,
Phys.\ Rev.\ D {\bf 40}, 1753 (1989);
N.~Makino and M.~Sasaki,
Prog.\ Theor.\ Phys.\  {\bf 86}, 103 (1991);
D.~I.~Kaiser,
Phys.\ Rev.\ D {\bf 52}, 4295 (1995).

\bibitem{Gumjudpai}
E.~Komatsu and T.~Futamase,
Phys.\ Rev.\ D {\bf 59}, 064029 (1999);
S.~Tsujikawa and B.~Gumjudpai,
Phys.\ Rev.\ D {\bf 69}, 123523 (2004).
  
\bibitem{unitarity} 
C.~P.~Burgess, H.~M.~Lee and M.~Trott,
JHEP {\bf 0909}, 103 (2009);
J.~L.~F.~Barbon and J.~R.~Espinosa,
Phys.\ Rev.\ D {\bf 79}, 081302 (2009);
C.~P.~Burgess, H.~M.~Lee and M.~Trott,
JHEP {\bf 1007}, 007 (2010);
R.~N.~Lerner and J.~McDonald,
JCAP {\bf 1004}, 015 (2010).
  
\bibitem{Germani:2010gm} 
C.~Germani and A.~Kehagias,
Phys.\ Rev.\ Lett.\  {\bf 105}, 011302 (2010);
C.~Germani and A.~Kehagias,
JCAP {\bf 1005}, 019 (2010).
  
\bibitem{Amendola:1993uh} 
L.~Amendola,
Phys.\ Lett.\ B {\bf 301}, 175 (1993).
  
\bibitem{Tsujikawa:2012mk} 
S.~Tsujikawa,
Phys.\ Rev.\ D {\bf 85}, 083518 (2012).

\bibitem{slotheon} 
C.~Germani, L.~Martucci and P.~Moyassari,
Phys.\ Rev.\ D {\bf 85}, 103501 (2012).

\bibitem{GermaniNa} 
C.~Germani and A.~Kehagias,
Phys.\ Rev.\ Lett.\  {\bf 106}, 161302 (2011).

\bibitem{Freese} 
K.~Freese, J.~A.~Frieman and A.~V.~Olinto,
Phys.\ Rev.\ Lett.\  {\bf 65}, 3233 (1990);
F.~C.~Adams, J.~R.~Bond, K.~Freese, J.~A.~Frieman and A.~V.~Olinto,
Phys.\ Rev.\ D {\bf 47}, 426 (1993).

\bibitem{Savage:2006tr} 
C.~Savage, K.~Freese and W.~H.~Kinney,
Phys.\ Rev.\ D {\bf 74}, 123511 (2006).
  
\bibitem{Banks:2003sx} 
T.~Banks, M.~Dine, P.~J.~Fox and E.~Gorbatov,
JCAP {\bf 0306}, 001 (2003);  
N.~Barnaby and M.~Peloso,
Phys.\ Rev.\ Lett.\  {\bf 106}, 181301 (2011).

\bibitem{Watanabe} 
C.~Germani and Y.~Watanabe,
JCAP {\bf 1107}, 031 (2011).
  
\bibitem{Nicolis:2008in} 
A.~Nicolis, R.~Rattazzi and E.~Trincherini,
Phys.\ Rev.\ D {\bf 79}, 064036 (2009).
  
\bibitem{Deffayet:2009wt} 
C.~Deffayet, G.~Esposito-Farese and A.~Vikman,
Phys.\ Rev.\ D {\bf 79}, 084003 (2009);  
C.~Deffayet, S.~Deser and G.~Esposito-Farese,
Phys.\ Rev.\ D {\bf 80}, 064015 (2009).

\bibitem{deRham} 
C.~de Rham and A.~J.~Tolley,
JCAP {\bf 1005}, 015 (2010);
K.~Van Acoleyen and J.~Van Doorsselaere,
Phys.\ Rev.\ D {\bf 83}, 084025 (2011).

\bibitem{Kamada:2010qe} 
K.~Kamada, T.~Kobayashi, M.~Yamaguchi and J.~'i.~Yokoyama,
Phys.\ Rev.\ D {\bf 83}, 083515 (2011).
  
\bibitem{DPSV}     
C.~Deffayet, O.~Pujolas, I.~Sawicki and A.~Vikman,
JCAP {\bf 1010}, 026 (2010).  
  
\bibitem{G-de}   
F.~Silva and K.~Koyama,
Phys.\ Rev.\ D {\bf 80}, 121301 (2009);
T.~Kobayashi, H.~Tashiro and D.~Suzuki,
Phys.\ Rev.\ D {\bf 81}, 063513 (2010);
A.~De Felice and S.~Tsujikawa,
JCAP {\bf 1007}, 024 (2010);
R.~Gannouji and M.~Sami,
Phys.\ Rev.\ D {\bf 82}, 024011 (2010).
  
\bibitem{G-inf} 
T.~Kobayashi, M.~Yamaguchi and J.~Yokoyama,
Phys.\ Rev.\ Lett.\ \textbf{105}, 231302 (2010); 
C.~Burrage, C.~de Rham, D.~Seery and A.~J.~Tolley, 
JCAP \textbf{1101}, 014 (2011); 
S.~Mizuno and K.~Koyama, 
Phys.\ Rev.\ \textbf{D82}, 103518 (2010); 
A.~De Felice and S.~Tsujikawa,
Phys.\ Rev.\ Lett.\  {\bf 105}, 111301 (2010);
Phys.\ Rev.\ D {\bf 84}, 124029 (2011);
A.~De Felice, R.~Kase and S.~Tsujikawa,
Phys.\ Rev.\ D {\bf 83}, 043515 (2011);
P.~Creminelli {\it et al.,}
JCAP \textbf{1102}, 006 (2011). 
A.~Naruko and M.~Sasaki, 
Class.\ Quant.\ Grav.\ \textbf{28}, 072001 (2011); 
X.~Gao,
JCAP {\bf 1110}, 021 (2011);
T.~Kobayashi, M.~Yamaguchi and J.~Yokoyama,
Phys.\ Rev.\ \textbf{D83}, 103524 (2011);
S.~Renaux-Petel,
Class.\ Quant.\ Grav.\  {\bf 28}, 182001 (2011)
[Erratum-ibid.\  {\bf 28}, 249601 (2011)];
S.~Renaux-Petel, S.~Mizuno and K.~Koyama,
JCAP {\bf 1111}, 042 (2011).

\bibitem{quantum} 
M.~A.~Luty, M.~Porrati and R.~Rattazzi,
JHEP {\bf 0309}, 029 (2003);
K.~Hinterbichler, M.~Trodden and D.~Wesley,
Phys.\ Rev.\ D {\bf 82}, 124018 (2010);
G.~Goon, K.~Hinterbichler and M.~Trodden,
JCAP {\bf 1107}, 017 (2011).

\bibitem{Popa} 
L.~A.~Popa,
JCAP {\bf 1110}, 025 (2011).

\bibitem{Kobayashi:2011nu} 
T.~Kobayashi, M.~Yamaguchi and J.~'i.~Yokoyama,
Prog.\ Theor.\ Phys.\  {\bf 126}, 511 (2011).

\bibitem{Gao} 
X.~Gao and D.~A.~Steer,
JCAP {\bf 1112}, 019 (2011).
 
\bibitem{DeFelice11}  
A.~De Felice and S.~Tsujikawa,
Phys.\ Rev.\ D {\bf 84}, 083504 (2011).

\bibitem{DeFelice:2011bh} 
A.~De Felice and S.~Tsujikawa,
JCAP {\bf 1202}, 007 (2012).

\bibitem{Horndeski}
G. W. Horndeski, Int.\ J.\ Theor.\ Phys. {\bf 10}, 363 (1974).

\bibitem{Deffayet11} 
C.~Deffayet, X.~Gao, D.~A.~Steer and G.~Zahariade,
Phys.\ Rev.\ D {\bf 84}, 064039 (2011).

\bibitem{Char} 
C.~Charmousis, E.~J.~Copeland, A.~Padilla and P.~M.~Saffin,
Phys.\ Rev.\ Lett.\  {\bf 108}, 051101 (2012).

\bibitem{DKT11} 
A.~De Felice, T.~Kobayashi and S.~Tsujikawa,
Phys.\ Lett.\ B {\bf 706}, 123 (2011).

\bibitem{Kamada2} 
K.~Kamada, T.~Kobayashi, T.~Takahashi, 
M.~Yamaguchi and J.~'i.~Yokoyama,
arXiv:1203.4059 [hep-ph].
 
\bibitem{Co_Per} 
J.~M.~Bardeen,
Phys.\ Rev.\ D {\bf 22}, 1882 (1980); 
H.~Kodama and M.~Sasaki,
Prog.\ Theor.\ Phys.\ Suppl.\  {\bf 78}, 1 (1984); 
V.~F.~Mukhanov, H.~A.~Feldman and R.~H.~Brandenberger,
Phys.\ Rept.\  {\bf 215}, 203 (1992); 
K.~A.~Malik and D.~Wands,
Phys.\ Rept.\  {\bf 475}, 1 (2009). 
      
\bibitem{Toporensky}      
S.~Alexeyev, A.~Toporensky and V.~Ustiansky,
Phys.\ Lett.\ B {\bf 509}, 151 (2001);     
A.~Toporensky and S.~Tsujikawa,
Phys.\ Rev.\ D {\bf 65}, 123509 (2002).       
    
\bibitem{super1} 
 A.~Adams, N.~Arkani-Hamed, S.~Dubovsky, A.~Nicolis and R.~Rattazzi,
 JHEP {\bf 0610}, 014 (2006);
C.~Bonvin, C.~Caprini and R.~Durrer,
Phys.\ Rev.\ Lett.\  {\bf 97}, 081303 (2006);
G.~Ellis, R.~Maartens and M.~A.~H.~MacCallum,
Gen.\ Rel.\ Grav.\  {\bf 39}, 1651 (2007).
  
\bibitem{super2}
E.~Babichev, V.~Mukhanov and A.~Vikman,
JHEP {\bf 0802}, 101 (2008).

\bibitem{super2d}
J.~P.~Bruneton,
Phys.\ Rev.\  D {\bf 75}, 085013 (2007);
R.~Geroch,
arXiv:1005.1614 [gr-qc].

\bibitem{super3}
J.~Evslin and T.~Qiu,
JHEP {\bf 1111}, 032 (2011);
J.~Evslin,
JHEP {\bf 1203}, 009 (2012).

\bibitem{super4}
C.~Burrage, C.~de Rham, L.~Heisenberg and A.~J.~Tolley,
JCAP {\bf 1207}, 004 (2012).    

\bibitem{Hawking}
S.~W.~Hawking, 
Phys.\ Rev.\ D {\bf 46}, 603 (1992).       
      
\bibitem{Percival:2009xn} 
W.~J.~Percival {\it et al.}  [SDSS Collaboration],
Mon.\ Not.\ Roy.\ Astron.\ Soc.\  {\bf 401}, 2148 (2010).
  
\bibitem{Riess:2009pu} 
A.~G.~Riess  {\it et al.},
Astrophys.\ J.\  {\bf 699}, 539 (2009).

\bibitem{multi_natudal} 
 J.~E.~Kim, H.~P.~Nilles and M.~Peloso,
JCAP {\bf 0501}, 005 (2005);
S.~Dimopoulos, S.~Kachru, J.~McGreevy and J.~G.~Wacker,
JCAP {\bf 0808}, 003 (2008);
R.~Easther and L.~McAllister,
JCAP {\bf 0605}, 018 (2006);  
L.~McAllister, E.~Silverstein and A.~Westphal,
Phys.\ Rev.\ D {\bf 82}, 046003 (2010).
            
\bibitem{KLS} 
L.~Kofman, A.~D.~Linde and A.~A.~Starobinsky,
Phys.\ Rev.\ D {\bf 56}, 3258 (1997).
      
\bibitem{lattice} 
S.~Y.~Khlebnikov and I.~I.~Tkachev,
Phys.\ Rev.\ Lett.\  {\bf 77}, 219 (1996);
S.~Y.~Khlebnikov and I.~I.~Tkachev,
Phys.\ Rev.\ Lett.\  {\bf 79}, 1607 (1997);
T.~Prokopec and T.~G.~Roos,
Phys.\ Rev.\ D {\bf 55}, 3768 (1997);
G.~N.~Felder and I.~Tkachev,
Comput.\ Phys.\ Commun.\  {\bf 178}, 929 (2008).
         
\bibitem{Planck:2006aa} 
[Planck Collaboration],
astro-ph/0604069.
 


\end{thebibliography}
\end{document}